\documentclass[prx,twocolumn,amsmath,amssymb,superscriptaddress,floatfix,nofootinbib,aps]{revtex4-2}

\usepackage{amsmath}
\usepackage{amsfonts}
\usepackage{graphicx}
\usepackage{dcolumn}
\usepackage{bm}
\usepackage{amsthm}
\usepackage{algorithm}
\usepackage{algpseudocode}
\usepackage[cmyk]{xcolor}
\usepackage[colorlinks,bookmarks=false,citecolor=magenta,linkcolor=magenta,urlcolor=magenta]{hyperref}
\usepackage{braket}
\usepackage{orcidlink}
\def\tr{\mathrm{tr}}

\definecolor{navy}{cmyk}{0.9,0.4,0,0.1}

\begin{document}

\preprint{APS/123-QED}

\title{Stabilizer R\'enyi Entropy for Translation-Invariant Matrix Product States}

\author{Lei-Yi-Nan Liu \orcidlink{0009-0005-3072-3650}}
\affiliation{%
 School of Physics, Beihang University, Beijing 100191, China
}

\author{Su Yi}
\affiliation{Institute of Fundamental Physics and Quantum Technology \& School of Physical Science and Technology, Ningbo University, Ningbo, 315211, China}
\affiliation{Peng Huanwu Collaborative Center for Research and Education, Beihang University, Beijing 100191, China}

\author{Jian Cui \orcidlink{0000-0001-6643-7625}}
\email{jiancui@buaa.edu.cn}
\affiliation{%
 School of Physics, Beihang University, Beijing 100191, China
}%

\date{\today}

\begin{abstract}

Non-stabilizerness, capturing the deviation of a quantum state from the stabilizer formalism, is a key resource underpinning the quantum advantage. The recently introduced stabilizer Rényi entropy (SRE) offers a tractable measure of non-stabilizerness, avoiding the complexity of conventional methods.
We study SRE in translation-invariant matrix product states (MPS), deriving exact expressions for representative states and introducing a numerically stable algorithm, named bond-DMRG, to compute the SRE density in infinite systems. Applying this method, we obtain high-precision SRE densities for the ground state of the one-dimensional Ising model. 
We also analyze non-onsite SRE density, showing it is bounded by a universal function of entanglement entropy, and further prove that two-site mutual SRE vanishes asymptotically in injective MPS. 
Our work not only introduces a powerful method for extracting the SRE density in quantum many-body systems, but also numerically reveals a fundamental connection between non-stabilizerness and entanglement, thereby paving the way for deeper theoretical investigations into their interplay.

\end{abstract}

\maketitle

\section{Introduction}
Quantum entanglement, which captures the non-classical correlation between different degrees of freedom within a quantum state~\cite{Horodecki2009}, plays a fundamental role in unraveling collective phenomena. These range from quantum phase transitions \cite{Osborne2002}, superconductivity \cite{Vedral2004}, many-body localization \cite{Abanin2019} to the emergence of exotic phases, including (symmetry-protected) topological phases, spin liquids and other nontrivial quantum states \cite{Tantivasadakarn2024,Chen2010,XiaoGang2019,Broholm2020}.
Beyond fundamental physics, entanglement underpins quantum computation \cite{Bennett1998,Jozsa1997}, quantum error correction \cite{Gottesman2009}, and classical simulation methods based on tensor networks \cite{Vidal2003,Schollwock2011}.

However, entanglement alone does not ensure quantum advantage. By the Gottesman-Knill theorem, even highly entangled stabilizer states admit efficient classical simulation \cite{Gottesman1998}. The resource enabling one to go beyond this limitation is termed non-Cliffordness or non-stabilizerness \cite{Bravyi2005,Veitch2014}. 
Existing measures of non-stabilizerness are either computationally demanding, requiring optimization over all stabilizer decompositions, or lack a direct interpretation as expectation values of observables \cite{Veitch2014,Bravyi2016,Bravyi2019,Bravyi2016PRL,Pashayan2015,Howard2017,Liu2022}. In contrast, the stabilizer Rényi entropy (SRE) avoids such minimization \cite{Lorenzo2022} and offers a tractable alternative, though its exact evaluation still scales as $4^n$ Pauli strings, restricting applications to small systems.

The relationship between entanglement and non-stabilizerness is also interesting, as 
both of which are fundamental resources for quantum computation. 
Traditionally, these quantities have been regarded as distinct and largely uncorrelated. 
Stabilizer states can exhibit high entanglement despite having zero non-stabilizerness, while product states can possess significant non-stabilizerness without entanglement. This distinction arises because entanglement is invariant under local unitary transformations, whereas non-stabilizerness is sensitive to both local and non-local transformations. 
But recent work demonstrate that non-stabilizerness correlates strongly with the structure of entanglement \cite{Tirrito2024}, and that their interplay can induce Hilbert space fragmentation and computational phase transitions \cite{Andi2025}. 
This naturally raises the question: what remains if one removes the contribution from local unitaries and focuses solely on the non-local component of non-stabilizerness, and how does it relate to entanglement?

Going one step further, integrating entanglement and non-stabilizerness offers new avenues for quantum information processing and for probing many-body systems. Within matrix product states (MPS), the SRE can be computed in polynomial time \cite{Haug2023MPS,Tarabunga2024,Guglielmo2023,Guglielmo2024}, 
but studies so far are limited to finite systems. In the 1D Ising model, for example, 
the SRE density peak deviates from the critical point due to finite-size effects. Although finite-size scaling techniques can sometimes be used to argue that the SRE density peak converges to the critical point in the thermodynamic limit. However, such extrapolations are generally unreliable for more complex or less understood models. In fact, full-state SRE is not universally tied to criticality \cite{Tarabunga2023}, suggesting that finite-size analysis may obscure key physics. Moreover, larger systems entail prohibitive computational costs and reduced precision. These limitations motivate methods that directly access the thermodynamic limit, such as translationally invariant MPS (TIMPS), which naturally describe infinite 1D systems.

To answer the abovementioned questions, we develop a framework to compute the SRE in TIMPS, deriving analytical results for representative states such as spin-coherent, GHZ-like, W, and Dicke states \cite{Cabello2002,Hillery1999,Fleischhaue2002,Prevedel2009,Bartschi2019}. To overcome the limitations of finite-size methods, we extend the analysis to infinite MPS (iMPS) and introduce bond-DMRG, a stable algorithm for evaluating the SRE density with high precision. Applying this approach to the 1D Ising model, we show that the SRE density peaks at the critical point. We further explore the behavior of SRE and non-onsite SRE densities across different ensembles of random iMPS. Besides, we study the mutual SRE (mSRE) as a probe of long-range non-stabilizerness, proving that it vanishes for injective iMPS at long distances and demonstrate this in the Ising model, where the mSRE not only decays rapidly but also exhibits non-analytic behavior in its derivative at the critical point under unbroken $\mathbb{Z}_2$ symmetry.

\section{SRE for TIMPS}
The $\alpha$-stabilizer R\'enyi entropy ($\alpha$-SRE) for a $n$-qubit pure state $\ket{\psi}$ is defined as 
\begin{equation}
    M_{\alpha}(\ket{\psi}):=\frac{1}{1-\alpha}\log\sum_{P\in\mathcal{P}_n}\Xi_P^{\alpha}(\ket{\psi})-\log d \label{definition}
\end{equation}
where $\mathcal{P}_n$ is the group of all $n$-qubit Pauli strings with $+1$ phase, and 
$\Xi_P(\ket{\psi}):=d^{-1}\bra{\psi}P\ket{\psi}^2$ with $d=2^n$ the dimension of $n$-qubit Hilbert 
space \cite{Lorenzo2022}. 
In this work, we focus on 2-SRE, i.e., the case where $\alpha = 2$ as recent work has proved that these $\alpha\ge 2$ SRE are magic monotones~\cite{Haug2023monotone,Lorenzo2024}. 

\begin{figure*}
    \centering
    \includegraphics[width=\linewidth]{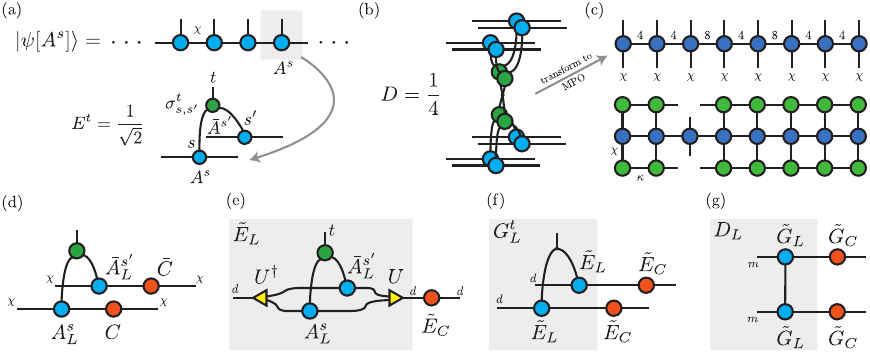}
    \caption{
    (a) and (b) show how tensors are contracted in Eq.~\eqref{transfer_matrix} and Eq.~\eqref{define_D} respectively. 
    (c) The basic idea of bond-DMRG algorithm is to construct the MPO form of matrix $D$ and perform standard DMRG to 
    obtain the dominant eigenvalue. The bond dimension of matrix $D$'s MPO is no more than 8 thus making bond-DRMG very efficient 
    when $\chi$ is not too large. 
    (d-g) show how tensors are contracted in the Pauli-iMPS algorithm. 
    \label{tensor-diagram}
    }
\end{figure*}

For an $n$-qubit quantum state $\ket{\psi}$ represented by a TIMPS characterized by a 
tensor $A^s$ with bond dimension $\chi$: 
$\ket{\psi[A^s]}=\sum_{s_1,\cdots,s_n}\mathrm{tr}(A^{s_1}\cdots A^{s_n})\ket{s_1\cdots s_n}$, 
we define four transfer matrices $E^t$ as shown in Fig.~\ref{tensor-diagram}(b)
\begin{equation}
    E^t = \sum_{s,s'=0}^{1}\frac{\sigma^t_{s,s'}}{\sqrt{2}}(A^s\otimes \bar{A}^{s'}), \label{transfer_matrix}
\end{equation}
where $t$ can be $0, 1, 2, 3$ and $\sigma^t_{s,s'}$ is the element of Pauli matrix $\sigma^t$.
The $\bar{A}^s$ is the complex conjugate of tensor $A^s$. 
Now, the quantity $\sum_{P\in\mathcal{P}_n}\Xi_P^{\alpha}(\ket{\psi})$ can be directly expressed using $E^t$, 
so the SRE for TIMPS is simply
\begin{equation}
    M_{\alpha}(\ket{\psi}) = \frac{1}{1-\alpha}\log \mathrm{tr}\left[\left(\sum_{t=0}^3 (E^t)^{\otimes 2\alpha}\right)^n\right]-n\log2. \label{SRE_for_TIMPS}
\end{equation}
Direct calculating $||\Xi(\ket{\psi})||^{2}$ in terms of $E^t$ is difficult because 
$E^t$ is a $\chi^2\times\chi^2$ matrix, and $(E^t)^{\otimes 4}$ has order $\chi^8$, which grows
rapidly. Therefore, truncation is required to calculate the SRE efficiently. 
Note that we are not restricted to a single tensor in the unit cell. The formalism can be straightforwardly generalized to multi-site unit cells.

\begin{figure}
    \includegraphics[width=\linewidth]{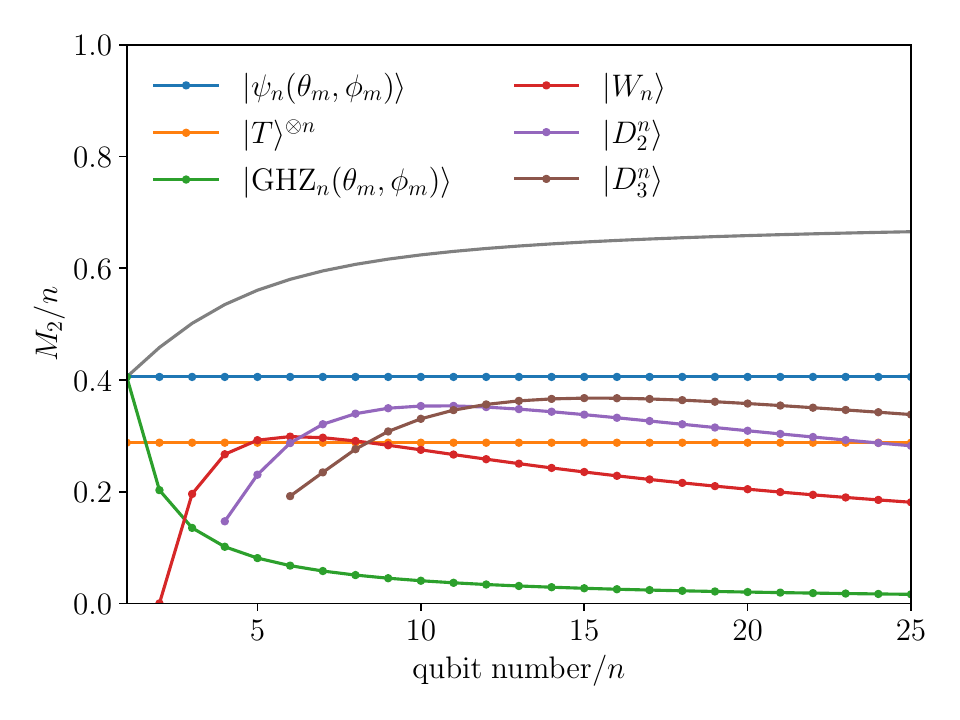}
    \caption{
    The figure displays the SRE densities of the states introduced in Sec.~I. The quantity $\mathrm{Ub}_n = [\log(2^n+1) - \log 2]/n$ represents the theoretical upper bound, while $\theta_m$ and $\phi_m$ denote the parameter values that maximize the SRE of $\ket{\psi_1(\theta,\phi)}$.
    \label{analytical_result}
    }
\end{figure}

Based on Eq.~\eqref{SRE_for_TIMPS}, some analytical 
results can be obatined for simple MPS. For example, the SRE of $n$-qubit W and Dicke states is 
\begin{eqnarray}
    M_2(\ket{D^n_1})&=&M_2(\ket{W_n})=3\log n-\log(7n-6), \\
    M_2(\ket{D^n_2})&=&3\log n(n-1) -2\log2 \nonumber\\ 
    &{}&-\log (91n^2-427n+492).
\end{eqnarray}
We also calculate SRE for spin-coherent, GHZ-like, and $\ket{D^n_3}$ states. 
The spin-coherent and GHZ-like states are defined as 
\begin{eqnarray}
    \ket{\psi_n(\theta,\phi)}&=&\left(\cos\frac{\theta}{2}\ket{0}+e^{i\phi}\sin\frac{\theta}{2}\ket{1}\right)^{\otimes n}, \\
    \ket{\mathrm{GHZ}_n(\theta,\phi)} &=& \cos\frac{\theta}{2}\ket{0}^{\otimes n} + e^{i\phi}\sin\frac{\theta}{2}\ket{1}^{\otimes n}. 
\end{eqnarray}
The details of these calculations are provided in \cite{SM}. 
As a summary, we show the SRE densities $m_2:=M_2/n$ of the 
discussed states and commmonly used magic state $\ket{T}^{\otimes n}=\ket{\psi_n(\pi/2,\pi/4)}$ 
in Fig.~\ref{analytical_result}. 

\section{SRE in infinite MPS}

Similar to other thermodynamic quantities, the SRE is an extensive property. 
We therefore investigate the density of SRE for an infinite MPS (iMPS), namely a TIMPS in the thermodynamic limit. 
The upper bound of 2-SRE $M_2$ directly provides the upper bound of $m_2$ in the thermodynamic limit \cite{Lorenzo2022}, i.e., 
$\lim_{n\to\infty}m_2(\ket{\psi})<\log2$. 
Here, we take an iMPS with a single tensor in the unit cell as an example. 
For a quantum state $\ket{\psi}$ represented by an iMPS 
characterized by a tensor $A^s$ with bond dimension $\chi$, 
we denote $D$ and its eigendecomposition 
\begin{equation}
    D=\sum_{t=0}^3(E^t)^{\otimes 4}=\sum_{i=0}^{\chi^8-1}\lambda_i(D)\ket{\lambda_i(D)}\bra{\lambda_i(D)}, \label{define_D}
\end{equation}
where $\lambda_i(D)$ is the $i$-th eigenvalue 
and we assume $|\lambda_0(D)|>|\lambda_1(D)|\ge\cdots\ge|\lambda_{\chi^8-1}(D)|$, 
since $D$ is generally non-Hermitian and its eigenvalues are therefore not necessarily real. 
Nevertheless, one can show that if the dominant eigenvalue $\lambda_0(D)$ is unique, then it must be real and positive as 
$\mathrm{tr}(D^n)>0$ for all $n$. 
Under this condition, the SRE density of $\ket{\psi}$ is given by 
\begin{equation}
    m_2(\ket{\psi}) = -\log \lambda_0(D)-\log 2,
\end{equation}
which has previously been obtained in \cite{Haug2023MPS,Ryan2024,David2025}. 

\begin{figure*}
    \centering
    \includegraphics[width=\linewidth]{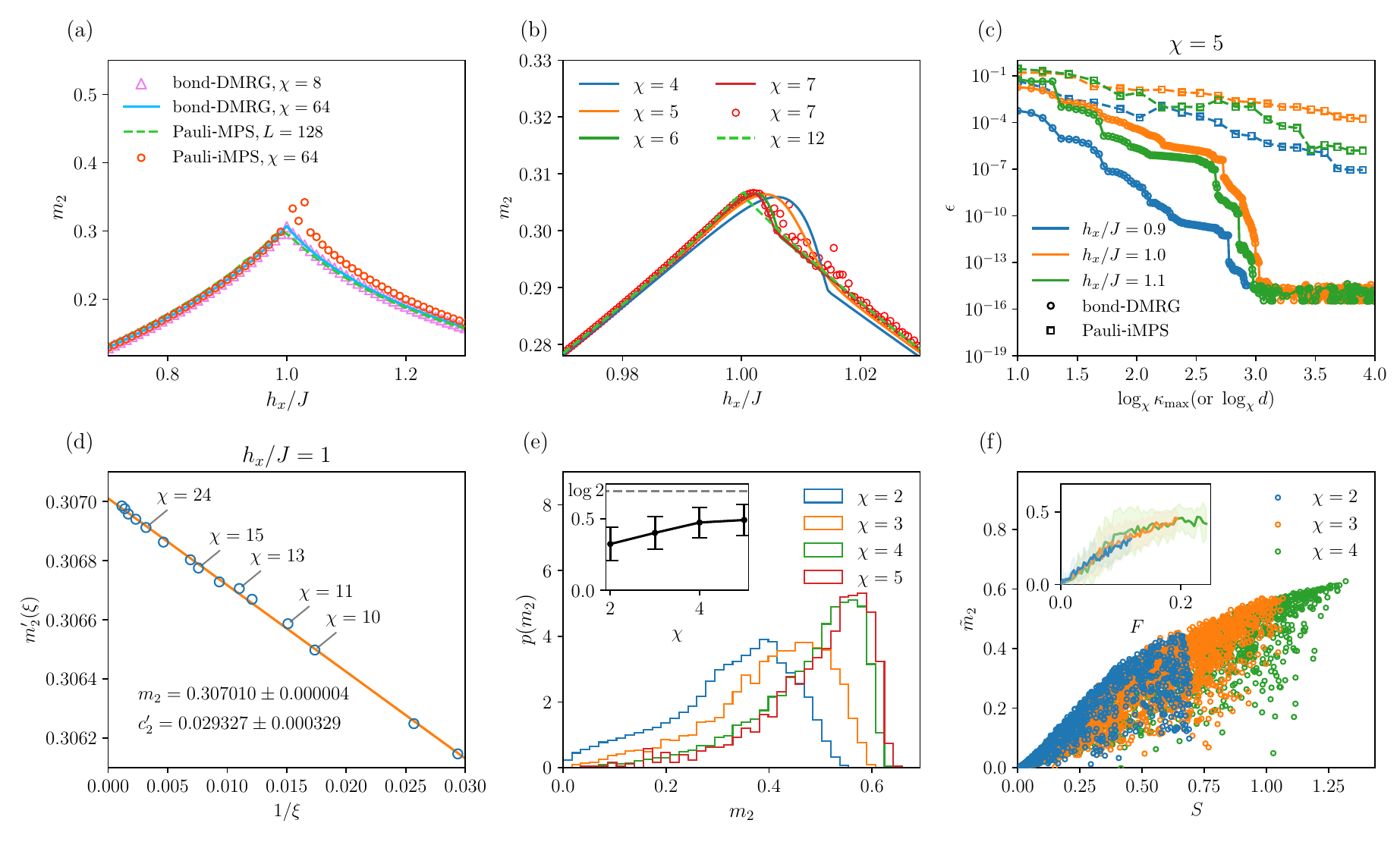}
    \caption{
    (a) Comparison of the SRE density $m_2$ obtained from bond-DMRG and Pauli-iMPS in the thermodynamic limit, together with finite-size Pauli-MPS results for $L=128$.
    (b) SRE density near the critical point obtained from Pauli-iMPS (solid lines), truncated Pauli-iMPS (red circles), and bond-DMRG (green dashed line).
    (c) Relative error $\epsilon$ of the two methods in different phases at bond dimension $\chi = 5$.
    (d) SRE density $m_2$ plotted as a function of $1/\xi$. Different values of the correlation length $\xi$ are generated by varying the bond dimension, revealing a clear dependence of $m_2$ on $\xi$.
    (e) The distributions of the SRE density $m_2$ for random iMPS. Inset: the average value of $m_2$ increases with the bond dimension $\chi$. 
    (f) The sampled data. The non-onsite SRE density $\tilde{m}_2$ appears to be bounded by a universal function of the entanglement entropy, i.e., $\tilde{m}_2 \leq f(S)$, independently of the bond dimension $\chi$. Inset: the relationship between $\tilde{m}_2$ and the logarithmic flatness $F$.
    \label{numerics}
    }
\end{figure*}

\begin{algorithm}[H]
\caption{Pauli iMPS algorithm}\label{PauliMPS_algorithm}
    \begin{algorithmic}[1]
    \Require local tensor of an iMPS $A^s$, cutoffs $\epsilon_1, \epsilon_2$ or max bond dimension $d, m$
    \State turn $A^s$ into left canonical form $A^s \to \{A_L^s, C\}$
    \State calculate $E_L^t$ using Eq.~\eqref{transfer_matrix}, and let $E_C=C\otimes\bar{C}$
    \State perform SVD and truncate singular values smaller than $\epsilon_1$ such that 
    $E_C\approx U_{\chi^2\times d}S_{d\times d}V^{\dagger}_{d\times \chi}$
    \State $\tilde{E}_L^t\gets U^{\dagger} E_L^t U$, $\tilde{E}_C\gets S_{d\times d}$ \Comment{$d\le\chi^2$, see Fig.~\ref{tensor-diagram}(e)}
    \State $G^t_L\gets \sum_{t_1,t_2}\delta^{t}_{t_1,t_2}(\tilde{E}^{t_1}_{L}\otimes \tilde{E}^{t_2}_L)$, $G_C\gets \tilde{E}_C^{\otimes 2}$
    \State perform SVD and truncate singular values smaller than $\epsilon_2$ such that 
    $G_C\approx U_{d^2\times m}S_{m\times m}V^{\dagger}_{m\times d^2}$
    \State $\tilde{G}_L^t\gets U^{\dagger}G^t_L U$, $\tilde{G}_C\gets S_{m\times m}$ \Comment{$m\le d^2\le \chi^4$}
    \State $D_L\gets \sum_t\tilde{G}_L^t\otimes\tilde{G}_L^t$
    \State calculate the dominant eigenvalue of $D_L$ and the SRE density is obtained
    \end{algorithmic}
\end{algorithm}

Due to the prohibitively large bond dimension of $D$, it is essential to reduce 
its bond dimension prior to computing the its dominant eigenvalue. 
A natural strategy to achieve this is to adopt 
the Pauli-MPS method proposed in \cite{Tarabunga2024}, adapted to the iMPS setting. 
We refer to this modified approach as the Pauli-iMPS method and we list the core steps of 
this method in Algorithm \ref{PauliMPS_algorithm}. The detailed procedures are provided in \cite{SM}.

Here we briefly outline the Pauli-iMPS method. 
To efficiently reduce the bond dimension of $D$, we do not construct $D=\sum_{t=0}^{3}(E^t)^{\otimes 4}$ 
in a single step. Instead, we first 
truncate the bond dimension of $E_L^t$ by discarding the small singular values of $E_C=C\otimes \bar{C}$. 
Specifically, we retain the largest $d$ singular values and use the corresponding unitary operator $U_{\chi^2\times d}$ to project $E_L^t$ onto a lower-dimensional subspace. 
As a result, the bond dimension is reduced from $\chi^2$ to $d$.
The truncated tensors are denoted by $\tilde{E}_L$ and $\tilde{E}_C$, respectively. 
Next, we construct $G_L^t = \sum_{t_1,t_2}\delta^t_{t_1,t_2}(\tilde{E}_L^{t_1}\otimes\tilde{E}_L^{t_2})$, 
which can be viewed as forming $(E^t)^{\otimes 2}$. At this stage, the bond dimension increases to $d^2$. 
We then truncate the small singular values of $G_C=\tilde{E}_C\otimes \tilde{E}_C$ retaining the largest m singular values. 
Finally, to obtain $D_L\sim (E^t)^{\otimes 4}$, we construct $D_L = \sum_t (\tilde{G}_L^t)^{\otimes 2}$ 
which effectively corresponds to squaring $\tilde{G}_L^t$. 
Through this procedure, the bond dimension is reduced from $\chi^8$ to $m^2$. 
Within the Pauli-iMPS framework, Krylov subspace methods enable an accurate computation of the SRE density for $\chi \lesssim 10$ at a moderate computational cost, corresponding to $m \lesssim 10^4$. For larger bond dimensions $\chi \gtrsim 10$, truncations are needed so the results become approximate.

To assess the numerical performance and stability of the Pauli-iMPS method, we apply it to the quantum Ising model in the thermodynamic limit. The ground state is obtained using the VUMPS algorithm \cite{Zauner2018,Vanderstraeten2019} for the Hamiltonian
\begin{equation}
\hat{H}=-h_x\sum_i \hat{\sigma}_i^x - J\sum_i \hat{\sigma}_i^z \hat{\sigma}_{i+1}^z
\end{equation}
with $J=1$ fixed. 
However, as shown in Fig.~\ref{numerics}(a), the Pauli-iMPS method exhibits pronounced numerical instabilities near the critical point. We find that this issue originates from its high sensitivity to truncation. To illustrate this, we compute the SRE density for a $\chi=7$ iMPS both with and without truncation. As shown in Fig.~\ref{numerics}(b), the untruncated result, obtained with $(d,m)=(\chi^2,\chi^4)=(49,2401)$, is smooth, although slightly affected by the small bond dimension. In contrast, introducing even a modest truncation, e.g. $(d,m)=(49,1600)$, leads to strongly jagged behavior. This demonstrates that small truncation errors can significantly distort the final result within the Pauli-iMPS framework.

To overcome this limitation, we develop a new algorithm termed bond-DMRG. 
The central idea is to construct a matrix product operator (MPO) representation of the matrix $D$ and determine its dominant eigenvalue $\lambda_0(D)$ using a DMRG-based variational procedure \cite{SRWhite1992,itensor}.
The construction of the MPO for $D$ is detailed in \cite{SM}.
Here we briefly outline the core idea of constructing the MPO representation of $D$.
The matrix $D$ is built from the iMPS tensor $A^s$ and its conjugate. Although $D$ formally acts on a very large Hilbert space, its structure consists of identical local building blocks connected by the physical  bonds of the matrix product density operator, as shown in Fig.~\ref{tensor-diagram}(b). 
By unfolding the underlying tree-like contraction structure into an one-dimensional chain, the operator $D$ can be reorganized as a sequence of local tensors connected along virtual bonds. In this way, the matrix $D$ is encoded in an MPO, allowing us to determine its dominant eigenvalue using standard MPO-based DMRG techniques without explicitly constructing the full matrix. 
Since the MPO for $D$ is generally non-Hermitian, the goal is not to minimize an energy functional, as in conventional ground-state DMRG, but to variationally target the eigenvalue of largest magnitude. Therefore, in the local optimization steps we employ an Arnoldi-type eigensolver, which is suitable for non-Hermitian operators and enables us to directly compute the dominant eigenvalue. 
The bond-DMRG results are presented in Fig.~\ref{numerics}(a) and (b). As a reference, Fig.~\ref{numerics}(a) also includes finite-size ($L=128$) results obtained via the 
Pauli-MPS method. While Pauli-iMPS becomes unstable near criticality, particularly for $h_x/J \ge 1$, 
bond-DMRG remains numerically stable across all parameters and agrees well with the baseline calculation.

To systematically compare the two approaches, we evaluate the SRE density for $\chi=5$ iMPS 
under various truncation parameters, taking the untruncated Pauli-iMPS result as the 
reference value $m_2^{\mathrm{ref}}$. We define the relative error as 
\begin{equation}
    \epsilon=\frac{|m_2^{\mathrm{cal}}-m_2^{\mathrm{ref}}|}{m_2^{\mathrm{ref}}}, 
\end{equation}
where $m_2^{\mathrm{cal}}$ denotes the truncated result. The errors for both methods are shown in Fig.~\ref{numerics}(c).
For bond-DMRG, the only truncation parameter is the maximum bond dimension of 
the MPS $\kappa_{\mathrm{max}}$, and we plot the error versus $\log_{\chi}\kappa_{\mathrm{max}}$. 
When $\kappa_{\mathrm{max}}=\chi^4$, no truncation is imposed and the error should vanish. 
In contrast, Pauli-iMPS involves two truncation parameters $(d,m)$. 
We consider data points satisfying $\log_{\chi} d \approx \log_d m$ and plot the error versus $\log_{\chi} m$. 
When $\log_{\chi} m=4$, the method becomes effectively untruncated.
As seen in Fig.~\ref{numerics}(c), bond-DMRG consistently achieves smaller relative errors 
across all parameter regimes. Once $\kappa_{\mathrm{max}} \gtrsim \chi^3$, the remaining error is dominated by numerical noise. Additional performance analysis are provided in \cite{SM}.

With the improved numerical stability of bond-DMRG, we are able to investigate the scaling relation 
between the SRE density $m_2$ and the correlation length $\xi$ at criticality. 
According to Ref.~\cite{Masahiro2025}, for a critical state $\ket{\psi}$ of a finite $L$-qubit system, 
the $\alpha$-SRE follows
\begin{equation}
    M_{\alpha}(\ket{\psi}) = m_{\alpha}L - c_{\alpha} + o(1),
    \label{CFT_prediction}
\end{equation}
where the leading term $m_{\alpha}L$ is non-universal, while the subleading constant $c_{\alpha}$ is universal and encodes properties of the underlying low-energy field theory. For the Ising model, $c_2 = \ln\sqrt{2}$. 
Although Eq.~\eqref{CFT_prediction} applies to finite systems, 
it can be extended to the thermodynamic limit by introducing the SRE density and focusing on $\alpha=2$:
\begin{equation}
    m'_2 = \lim_{L\to+\infty}\frac{M_{2}(\ket{\psi})}{L} = m_2-\frac{c_2}{L},
\end{equation}
where $m_2$ is the thermodynamic-limit value. 
For iMPS, the system size $L$ is not well defined. Instead, the correlation length 
$\xi$ provides a natural effective length scale. An iMPS with correlation length $\xi$ can be 
viewed as describing the bulk of a finite system with effective size $L = O(\xi)$. Replacing $L$ by $\xi$ therefore leads to the scaling relation 
\begin{equation}
    m'_2(\xi) = m_2-\frac{c_2'}{\xi}, \label{scaling}
\end{equation}
where $m’_2(\xi)$ is the numerical estimate and $c’_2$ is a fitting parameter. This form is expected to hold for other R\'enyi indices $\alpha$ as well.
We perform bond-DMRG calculations at the Ising critical point with bond dimensions $\chi$ ranging from 8 to 40. As shown in Fig.~\ref{numerics}(d), the numerical data follow the scaling behavior in Eq.\eqref{scaling}. Extrapolating to $\xi \to \infty$, we obtain a state-of-the-art estimate for the 2-SRE density of the critical Ising chain, $m_2 = 0.3070$.

Having established the stability and accuracy of the bond-DMRG method, 
we next use it to explore the relationship between SRE density and entanglement in iMPS 
characterized by random tensors of the form $(A^s)_{ij} = r^s_{ij} e^{i 2\pi \theta^s_{ij}}$, 
where $r^s_{ij}$ and $\theta^s_{ij}$ are independent random variables uniformly distributed in the interval $[0,1)$. 
The distribution of the SRE density $m_2$ for random iMPS is presented in Fig.~\ref{numerics} (e). 
As illustrated in the inset, the average value of $m_2$ increases with the bond dimension $\chi$. 
We have two notable observations. First, weakly entangled quantum states can still 
exhibit relatively high SRE density. 
Second, the distributions for different values of $\chi$ 
appear to converge toward a universal shape as $\chi$ increases. 
Determining the form of this distribution in the $\chi \to \infty$ limit presents 
an interesting direction for future research.

We now investigate the relationship between the SRE density $m_2$ and the entanglement spectrum $\{ p_i=\lambda_i^2 \}$ 
in random iMPS, for which both quantities can now be efficiently computed. 
To characterize the entanglement spectrum, we focus on two key properties. 
The first is the entanglement entropy, defined as $S = -\sum_ip_i\log p_i$. 
The second property is the (anti-)flatness of the entanglement spectrum, as introduced in \cite{Tirrito2024}. 
Consider a pure state $\ket{\psi}$ in a bipartite system $\mathcal{H}=\mathcal{H}_A\otimes \mathcal{H}_B$, with 
reduced density matrix $\rho_A = \mathrm{tr}_B |\psi\rangle\langle\psi|$. The (anti-)flatness is defined by
$\mathcal{F}_A(\ket{\psi}):=\mathrm{tr}(\rho_A^3) - [\mathrm{tr}(\rho_A^2)]^2$, 
which captures the deviation of the entanglement spectrum from uniformity. In Ref.~\cite{Tirrito2024}, 
it was further demonstrated that the flatness is related to the stabilizer linear 
entropy $M_{\mathrm{lin}}$. 
However, the stabilizer linear entropy is not well-defined in the thermodynamic limit. 
Specifically, one finds that its density $m_{\mathrm{lin}} = 0$ in the thermodynamic limit for all states. 
Therefore, we still focus on $m_2$ and study its relation with not flatness but ``logarithmic (anti-)flatness'' 
first introduced in \cite{Odavic2025}, which is defined as 
\begin{equation}
    F(\rho_A):=\log[\mathrm{tr}(\rho_A^3)] - \log[\mathrm{tr}(\rho_A^2)]^2.
\end{equation}

\begin{figure*}
    \includegraphics[scale=0.46]{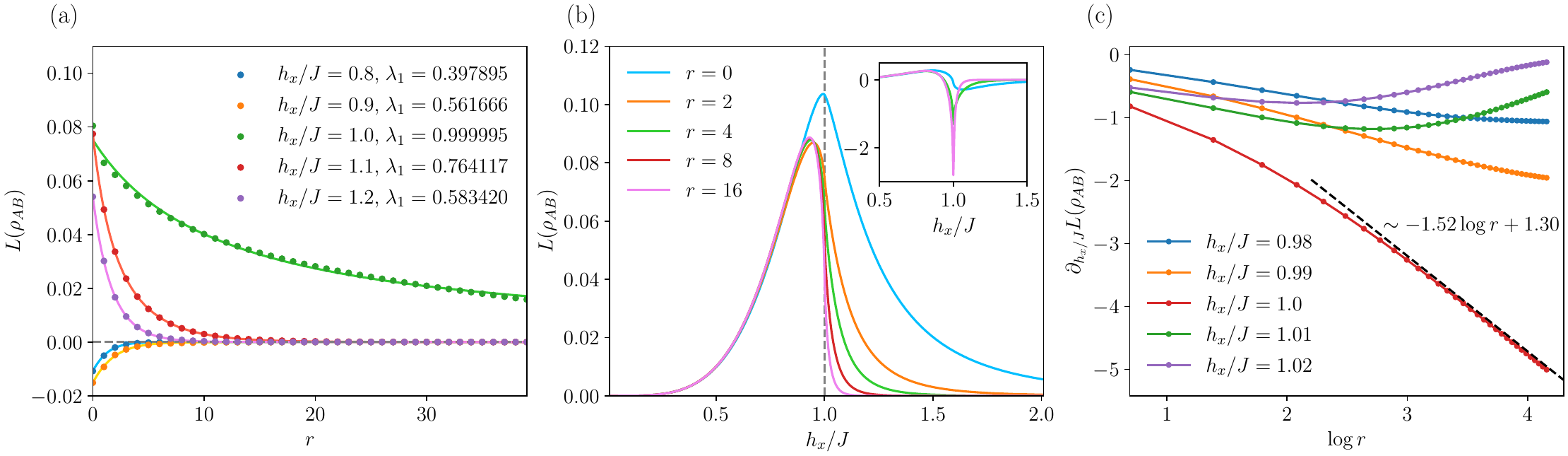}
    \caption{(a) The two-point mSRE in the ground state of the quantum Ising model. The colored dots represent the results from DMRG, while the lines are the fitting results. 
    (b) The exact mSRE $L(\rho_{AB})$ for the Ising model with unbroken $\mathbb{Z}_2$ symmetry shows a non-vanishing behaviour in the ferromagnetic phase and vanishing one in the paramagnetic phase. Inset: the derivative of $L(\rho_{AB})$ exhibits a sharp peak at the critical point as the distance between two subsystems increases. 
    (c) Derivative of $L(\rho_{AB})$ for the Ising model with unbroken $\mathbb{Z}_2$ symmetry near the critical point. It can be observed that the derivative at the critical point follows a logarithmic scaling behavior, as described in Eq.~\eqref{dL_scaling}. 
    \label{Ising_mSRE}
    }
\end{figure*}

For product states, the flatness vanishes, 
yet such states may still possess SRE. To capture the component of SRE that cannot 
be eliminated by on-site unitary transformations, we define the non-onsite SRE density for a $n$-quabit state as
\begin{equation}
    \tilde{m}_2(\ket{\psi}) = \min_{\{U_i\}} m_2\left[\left(\bigotimes_{i=1}^n U_i\right) \ket{\psi}\right], 
\end{equation}
where the minimization is performed over all single-site unitary operations $U_i\in \mathrm{SU(2)}$. 
This procedure removes contributions arising solely from local on-site unitary freedom, but it does not, in general, eliminate more structured multi-site correlations that may also be regarded as local in a broader sense. 
In finite systems, evaluating $\tilde{m}_2$ is computationally challenging, 
as the optimal local unitary can, in general, vary from site to site. However, for translationally 
invariant iMPS, this difficulty is mitigated: the optimal local unitary minimizing $m_2$ 
must be the same at every site. This symmetry greatly simplifies the 
minimization procedure and allows us to systematically explore the behavior of 
$\tilde{m}_2$ for iMPS.

We perform numerical evaluations of $\tilde{m}_2$ across ensembles of random iMPS to 
investigate its dependence on the entanglement structure. 
As shown in Fig.~\ref{numerics} (f), the non-onsite SRE density $\tilde{m}_2$ 
are observed to be upper bounded by a function of the entropy $S$, i.e., 
$\tilde{m}_2 \le f(S)$, where the function $f$ appears to be independent of the bond dimension $\chi$. 
This indicates there is a universal constraint on the maximal amount of non-onsite SRE that a state can support for a given level of EE and also suggests that non-onsite SRE provides a finer characterization of quantum correlations than EE. 
Furthermore, as shown in the inset of Fig.~\ref{numerics} (f), $\tilde{m}_2$ exhibits a clear positive correlation with the $F$ of the entanglement spectrum. 
This suggests that logarithmic flatness 
can serve as a useful indicator of non-onsite non-stabilizer resource. 
Taken together, these results demonstrate that both total and non-onsite SRE densities 
are closely related to the entanglement structure of the state, with $\tilde{m}_2$ 
providing a more refined and locally robust measure of non-stabilizerness.

\section{mutual SRE for TIMPS}

We first introduce the $2$-SRE definition for a $n$-qubit mixed state $\rho$: 
$\tilde{M}_2(\rho):=M_2(\rho)-S_2(\rho)$, 
where ${M}_2(\rho)=-\log\sum_{P\in\mathcal{P}_n}\mathrm{tr}(P\rho)^4$ is the generalization of Eq.~\eqref{definition}
and $S_2(\rho) = -\log\mathrm{tr}(\rho^2)$ is the R\'enyi-2 entropy \cite{Lorenzo2022}. The 2-SRE 
for mixed state $\rho$ can be seen as the R\'enyi-2 entropy of 
$\tilde{\Xi}_P=|\mathrm{tr}(P\rho)^2|/\sum_{P\in\mathcal{P}_n}|\mathrm{tr}(P\rho)|^2$ apart from some offset 
and the free states are defined as the mixed states that can be obtained from pure stabilizer states by partial tracing \cite{Lorenzo2022,Tarabunga2023}. 

We can now define a measure of long-range non-stabilizerness, called mutual SRE (mSRE), 
analogous to mutual information, as 
\begin{equation}
    L(\rho_{AB})=\tilde{M}_2(\rho_{AB})-\tilde{M}_2(\rho_A)-\tilde{M}_2(\rho_B), \label{mSRE_definition}
\end{equation}
where $A$ and $B$ are two separated subsystems. This mSRE measures the extent to 
which non-stabilizerness is encoded in the correlations between subsystems, 
thereby quantifying the degree to which non-stabilizerness cannot be eliminated 
by finite-depth quantum circuits \cite{Tarabunga2023,White2021}. 
For an iMPS characterized by a tensor $A^s$, we consider two subsystems $A$ and $B$ separated 
by a distance $r$, where $s_2-s_1-1=r$ with $s_1$ and $s_2$ representing the sites of subsystems $A$ and 
$B$, respectively. Due to the translational invariance of iMPS, the mSRE $L(\rho_{AB})$ between A and B 
only depends on their distance $r$.

The mSRE consists of two parts $L(\rho_{AB})=L_{M}(\rho_{AB})+L_S(\rho_{AB})$ where $L_M(\rho_{AB})=M_2(\rho_{AB})-2M_2(\rho_A)$ and $L_S(\rho_{AB})=2S_2(\rho_A)-S_2(\rho_{AB})$. 
The $L_S(\rho_{AB})$ is actually the mutual information defined with R\'enyi-2 entropy. 
Denoting a new transfer matrix and its eigendecomposition
\begin{equation}
    E^{\bar{A}}_A=\sqrt{2}E^0=\sum_{s=0}^1 A^s\otimes \bar{A}^s
    =|r)(l|+\sum_i\lambda_i|\lambda_i)(\lambda_i|, 
\end{equation}
where the vectors $|l)$ and $|r)$ satisfy $(l|E^{\bar{A}}_A=(l|$ and 
$E^{\bar{A}}_A|r)=|r)$, we can directly obtain the expression for $\rho_A, \rho_B$ and $\rho_{AB}$
\begin{eqnarray}
    \rho_A &=& (l|(A^{s_1}\otimes \bar{A}^{s_1'})|r), \\
    \rho_B &=& (l|(A^{s_2}\otimes \bar{A}^{s_2'})|r), \\
    \rho_{AB}&=&(l|(A^{s_1}\otimes \bar{A}^{s_1'})(E^{\bar{A}}_A)^r(A^{s_2}\otimes \bar{A}^{s_2'})|r)\nonumber\\
    &=&\rho_A\otimes\rho_B+\sum_i\lambda_i^r\rho_l(\lambda_i)\otimes\rho_r(\lambda_i), 
\end{eqnarray}
where $\rho_l(\lambda_i)=(l|(A^{s_1}\otimes \bar{A}^{s_1'})|\lambda_i)$ and 
$\rho_r(\lambda_i)=(\lambda_i|(A^{s_2}\otimes \bar{A}^{s_2'})|r)$. 
By substituting the expressions of 
$\rho_A,\rho_B$ and $\rho_{AB}$ in Eq.~\eqref{mSRE_definition}, we obatin the long-range limit ($r\to+\infty$) for 
mSRE
\begin{equation}
    \lim_{r\to+\infty}L(\rho_{AB})=\lim_{r\to+\infty}\log\left[
    \frac{1+2\lambda_1^rc_s}{1+2\lambda_1^rc_m}\right] = 0, 
    \label{Lab}
\end{equation}
where $|\lambda_1|\in[0,1)$ is the second largest eigenvalue of transfer matrix $E^{\bar{A}}_A$ and $c_s,c_m$ are constants independent of $r$. The explicit expressions of $c_s$ and $c_m$ 
can be found in \cite{SM}. Therefore, we prove that for any injective iMPS, that is, an iMPS whose transfer matrix has one unique eigenvalue of largest magnitude \cite{Cirac2021}, the two-point mSRE will eventually decay to zero as the distance $r$ tends to infinity and the long-range behaviour of mSRE takes the form of Eq.~\eqref{Lab}. 

To verify our conclusion in Eq.~\eqref{Lab}, we perform DMRG \cite{SRWhite1992,itensor} calculation in a chain of size $L=128$ for Ising model and 
extract the mSRE between the site $s_1=64$ and $s_2=s_1+r+1$ with $r\in[0,39]$. 
Results are shown in Fig.~\ref{Ising_mSRE}(a) where we use the asymptotic behavior of $L(\rho_{AB})$ to extract the  $\lambda_1$.
It can be found that near the critical point $h_x/J=1$, $\lambda_1$ approches $1$ thus indicating the divergence of correlation length at $h_x/J=1$, as $\lambda_1$ is directly related to the correlation length with $\xi=-1/\ln|\lambda_1|$ \cite{Zauner2015, Vanderstraeten2019}. 

In addition to the decay behavior, the mSRE for the Ising model without symmetry breaking reveals a clear distinction: in the ferromagnetic phase, the mSRE remains finite, whereas in the paramagnetic phase, it vanishes, as shown in Fig.~\ref{Ising_mSRE}(b). This behavior can be attributed to the modified asymptotic form of $L(\rho_{AB})\sim\log \frac{1+2\lambda_1^rc_s}{1+2\lambda_1^rc_m}+\log c$ 
where the parameter $c$ arises due to the unbroken symmetry.
In Fig.~\ref{Ising_mSRE}(b), we also observe that the mSRE does not peak at the critical point. 
Instead, it peaks slightly before $h_x/J = 1$ for all values of $r$. 
However, the derivative of the mSRE exhibits a pronounced peak at the critical point, which becomes increasingly sharp with larger $r$, following a scaling behavior of
\begin{equation}
    \frac{\partial L(\rho_{AB})}{\partial (h_x/J)} \Bigg|_{h_x/J=1}\sim\log r, 
    \label{dL_scaling}
\end{equation}
as observed in Fig.~\ref{Ising_mSRE}(c). 
These observations lead to two conclusions. First, long-range entanglement appears to be a necessary condition for the emergence of long-range mSRE. Second, the derivative of the long-range mSRE may serve as a robust indicator of quantum phase transitions.

\section{Discussion and Outlook}
We investigate the SRE in TIMPS. After deriving analytical results, we develop an efficient iMPS algorithm and apply it to the 1D Ising model, where the SRE density peaks and becomes non-analytic at criticality. For random iMPS, we show that non-onsite SRE is bounded by a universal function of entanglement entropy and correlates with entanglement spectrum flatness. Finally, we find that the mutual SRE vanishes asymptotically in injective iMPS but displays critical non-analyticity in its derivative, indicating a connection to quantum phase transitions.

Several open questions remain for future investigation. 
It is of interest to determine the limiting form of the SRE density distribution of random iMPS as $\chi \to \infty$, 
and to identify the explicit form of the universal function bounding the non-onsite SRE. 
Furthermore, obtaining analytical expressions for the SRE and mSRE in the Ising model, 
particularly at the critical point using CFT techniques, would be highly valuable. 
In this context, recent progress has been made in relating SRE to CFT frameworks \cite{Masahiro2025}. 
Finally, investigating the connections between mSRE and other emerging measures, 
such as non-stabilizer entanglement entropy \cite{Mingpu2024} and quantum non-local non-stabilizerness \cite{Dongheng2025}, 
may provide deeper insights into the structure of non-stabilizerness in quantum many-body systems.

\begin{acknowledgments}
This work is supported by the National Key Research and Development 
of China (Grant Nos. 2021YFA1402001, 2021YFA0718304) and 
the National Natural Science Foundation of China (NSFC) (Grant Nos. 12375007, 12574295).
\end{acknowledgments}

\bibliography{main}

\newpage
\newpage
\appendix
\onecolumngrid
\section{SRE Calculation of Spin coherent, GHZ-like, W and Dicke states}\label{SRE_calculation}
\subsection{Spin coherent state}
For product state ($\chi=1$)
\begin{align}
    \ket{\psi_n(\theta,\phi)}=\left(\cos\frac{\theta}{2}\ket{0}+e^{i\phi}\sin\frac{\theta}{2}\ket{1}\right)^{\otimes n}, 
\end{align}
we can directly obtain 
\begin{align}
M_2(\ket{\psi_n(\theta,\phi)})=nM_2(\ket{\psi_1(\theta,\phi)}).
\end{align}
This is the result of the requirement from resource theory, i.e., the additivity of $\alpha$-SRE~\cite{Lorenzo2022}, 
\begin{align}
    M_{\alpha}(\ket{\psi}\otimes\ket{\phi})=M_{\alpha}(\ket{\psi})+M_{\alpha}(\ket{\phi}).
\end{align}

The SRE of a single spin coherent state $\ket{\psi_1(\theta,\phi)}$ is
\begin{align}
M_2(\ket{\psi_1(\theta,\phi)})=-\log\left[\cos^8\frac{\theta}{2}+\sin^8\frac{\theta}{2}+\frac{6+\cos 4\phi}{8}\sin^4\theta\right]
\end{align}
It reaches the maximum value $\log\frac{3}{2}$ at $(\theta_m,\phi_m)=(2\arccos\sqrt{\frac{3-\sqrt{3}}{6}},\frac{\pi}{4})$. 
Note that this is not the only maximum point. Due to the 
stability of SRE under Clifford operations \cite{Lorenzo2022}, the state $\ket{\psi_1(\theta_m,\phi_m)}$ with any Clifford gate acted on reaches the maximum value of SRE. 

\subsection{GHZ-like state}
The GHZ-like state, defined as 
\begin{equation}
    \ket{\mathrm{GHZ}_n(\theta,\phi)} = \cos\frac{\theta}{2}\ket{0}^{\otimes n} + e^{i\phi}\sin\frac{\theta}{2}\ket{1}^{\otimes n}, 
\end{equation}
with $\theta\in[0,\pi]$ and $\phi\in[0,2\pi)$, 
can be expressed as a simple MPS with bond dimension $\chi=2$. The site tensor $A^s$ for 
GHZ-like state is 
\begin{equation}
    A^0 = \begin{pmatrix}
        \sqrt[n]{\cos\frac{\theta}{2}} & 0\\
        0 & 0
    \end{pmatrix}, 
    A^1 = \begin{pmatrix}
        0 & 0\\
        0 & e^{i\phi/n}\sqrt[n]{\sin\frac{\theta}{2}}
    \end{pmatrix}.
\end{equation}

For TIMPS, we can calculate its SRE only using its local tensor. Here we can easily calculate the 
$E^t$ tensor for $\ket{\mathrm{GHZ}_n(\theta,\phi)}$ state
\begin{align}
    E^0 = \begin{pmatrix}
        aa^* &0 &0 &0 \\
        0& 0 & 0 &0 \\
        0& 0 & 0 &0 \\
        0& 0 & 0 & bb^*
    \end{pmatrix},
    E^1 = \begin{pmatrix}
        0 & 0 & 0 & 0 \\
        0 & ab^* & 0 & 0 \\
        0 & 0 & a^*b & 0 \\
        0 & 0 & 0 & 0
    \end{pmatrix}, 
    E^2=\begin{pmatrix}
        0 & 0 & 0 & 0 \\
        0 & -iab^* & 0 & 0 \\
        0 & 0 & ia^*b & 0 \\
        0 & 0 & 0 & 0
    \end{pmatrix},
    E^3 = \begin{pmatrix}
        aa^* & 0 & 0 & 0 \\
        0 & 0 & 0 & 0 \\
        0 & 0 & 0 & 0 \\
        0 & 0 & 0 & -bb^*
    \end{pmatrix},
\end{align}
where 
\begin{align}
    a &= \frac{1}{\sqrt{2}}\sqrt[n]{\cos \frac{\theta}{2}}, \\
    b &= \frac{e^{i\phi/n}}{\sqrt{2}}\sqrt[n]{\sin\frac{\theta}{2}}.
\end{align}
It is clear that all $E^t$ matrices are in diagonal form. 
Thus the following calculation is rather easy and we directly gives the SRE of the above state:
\begin{align}
    M_2(\ket{\mathrm{GHZ}_n(\theta,\phi)}) &= -\log\tr[ (E^{0\ \otimes 4}+E^{1\ \otimes 4}+E^{2\ \otimes 4}+E^{3\ \otimes 4})^n ]-n\log 2\\
    &= -\log\left[
        \cos^8\frac{\theta}{2}+\sin^8\frac{\theta}{2}+\frac{6+\cos4\phi}{8}\sin^4\theta
    \right]
\end{align}
which is equal to the expression of single qubit SRE. 
This is a natural result if we adopt the perspective of quantum circuit, as the 
GHZ-like state can be prepared by a single-qubit rotation gate followed by CNOT gates and CNOT gate 
is a Clifford operation so the non-stabilizerness can only be generated from the single-qubit rotation. 

\subsection{W state}
The W state is defined as 
\begin{align}
    \ket{W_n}=\frac{\ket{100\cdots0}+\ket{010\cdots0}+\ket{001\cdots0}+\cdots+\ket{000\cdots1}}{\sqrt{n}}.
\end{align}
The SRE of $n$-qubit W state has been calculated previously in \cite{Jovan2023,Catalano2025}, the result is 
\begin{align}
    M_2(\ket{W_n}) = 3\log n - \log(7n-6).
\end{align}
The previous paper directly calculated this result. Using TIMPS we can also obtain this result.

The MPS local tensor of $W$ state is 
\begin{align}
    L^0&=\begin{pmatrix}
        1 & 0
    \end{pmatrix}, 
    L^1=\begin{pmatrix}
        0 & \frac{1}{\sqrt{n}}
    \end{pmatrix},\\
    A^0&=\begin{pmatrix}
        1 & 0\\ 0 & 1
    \end{pmatrix},
    A^1 = \begin{pmatrix}
        0 & \frac{1}{\sqrt{n}}\\
        0 & 0
    \end{pmatrix},\\
    R^0&=\begin{pmatrix}
        0 \\ 1
    \end{pmatrix},
    R^1=\begin{pmatrix}
        \frac{1}{\sqrt{n}} \\ 0
    \end{pmatrix}, 
\end{align}
and the $E$ tensors are 
\begin{align}
    E_L^0&=\frac{1}{\sqrt{2}}\begin{pmatrix}
        1 & 0 & 0 & \frac1n
    \end{pmatrix}, 
    E_L^1=\frac{1}{\sqrt{2}}\begin{pmatrix}
        0 & \frac{1}{\sqrt{n}} & \frac{1}{\sqrt{n}} & 0
    \end{pmatrix},
    E_L^2=\frac{1}{\sqrt{2}}\begin{pmatrix}
        0 & -i\frac{1}{\sqrt{n}} & i\frac{1}{\sqrt{n}} & 0
    \end{pmatrix}, 
    E_L^3=\frac{1}{\sqrt{2}}\begin{pmatrix}
        1 & 0 & 0 & -\frac1n
    \end{pmatrix},\\
    E^0&=\frac{1}{\sqrt{2}}\begin{pmatrix}
        1 & 0 & 0 & \frac1n\\
        0 & 1 & 0 & 0\\
        0 & 0 & 1 & 0\\
        0 & 0 & 0 & 1
    \end{pmatrix},
    E^1=\frac{1}{\sqrt{2}}\begin{pmatrix}
        0 & \frac{1}{\sqrt{n}} & \frac{1}{\sqrt{n}} & 0\\
        0 & 0 & 0 & \frac{1}{\sqrt{n}}\\
        0 & 0 & 0 & \frac{1}{\sqrt{n}}\\
        0 & 0 & 0 & 0
    \end{pmatrix},
    E^2=\frac{1}{\sqrt{2}}\begin{pmatrix}
        0 & \frac{-i}{\sqrt{n}} & \frac{-i}{\sqrt{n}} & 0\\
        0 & 0 & 0 & \frac{-i}{\sqrt{n}}\\
        0 & 0 & 0 & \frac{-i}{\sqrt{n}}\\
        0 & 0 & 0 & 0
    \end{pmatrix},
    E^3=\frac{1}{\sqrt{2}}\begin{pmatrix}
        1 & 0 & 0 & -\frac1n\\
        0 & 1 & 0 & 0\\
        0 & 0 & 1 & 0\\
        0 & 0 & 0 & 1
    \end{pmatrix},\\
    E_R^0&=\frac{1}{\sqrt{2}}\begin{pmatrix}
        \frac1n \\ 0 \\ 0 \\ 1
    \end{pmatrix}, 
    E_R^1=\frac{1}{\sqrt{2}}\begin{pmatrix}
        0 \\ \frac{1}{\sqrt{n}} \\ \frac{1}{\sqrt{n}} \\ 0
    \end{pmatrix},
    E_R^2=\frac{1}{\sqrt{2}}\begin{pmatrix}
        0 \\ i\frac{1}{\sqrt{n}} \\ -i\frac{1}{\sqrt{n}} \\ 0
    \end{pmatrix}, 
    E_R^3=\frac{1}{\sqrt{2}}\begin{pmatrix}
        -\frac1n \\ 0 \\ 0 \\ 1
    \end{pmatrix},
\end{align}
so the SRE is 
\begin{align}
    M_2(\ket{W_n}) &= -\log\left[\left(\sum_{t=0}^3 E_L^{t\ \otimes 4}\right)\left(\sum_{t=0}^3 E^{t\ \otimes 4}\right)^{n-2}
    \left(\sum_{t=0}^3 E_R^{t\ \otimes 4}\right)\right]-n\log2\\
    &= 3\log n - \log(7n-6).
\end{align}

\subsection{Dicke states}
The Dicke state is defined as 
\begin{align}
    \ket{D^n_k}=\frac{1}{\sqrt{C_n^k}}\sum_i\mathcal{P}_i(\ket{0}^{\otimes n-k}\otimes\ket{1}^{\otimes k})
\end{align}
where $k=0,1,\cdots,n$ is the number of 1's, and the summation is over all distinct permutations, e.g., 
for $n=4,k=2$ we have
\begin{align}
    \ket{D_2^4}=\frac{\ket{1100}+\ket{1010}+\ket{1001}+\ket{0110}+\ket{0101}+\ket{0011}}{\sqrt{C_4^2}},
\end{align}
and the W state we discussed above is actually a special Dicke state with $k=1$,
\begin{align}
    \ket{W_n}=\ket{D_1^n}=\frac{\ket{100\cdots0}+\ket{010\cdots0}+\ket{001\cdots0}+\cdots+\ket{000\cdots1}}{\sqrt{C_n^1}}.
\end{align}

The Dicke state is a permutationally invariant state, therefore its SRE can be evaluated numerically in a more eﬀicient way \cite{Passarelli2024}. But the analytical expression of the Dicke state’s SRE is still not easy to obatin. 

For $k=2$ Dicke states, we have 
\begin{align}
    A_{L,1}^0&=\begin{pmatrix}
        1 & 0
    \end{pmatrix}, 
    A_{L,1}^1=\begin{pmatrix}
        0 & a
    \end{pmatrix},
    A_{L,2}^0=\begin{pmatrix}
        1 & 0 & 0\\
        0 & 1 & 0
    \end{pmatrix}, 
    A_{L,2}^1=\begin{pmatrix}
        0 & a & 0\\
        0 & 0 & 1
    \end{pmatrix}, \\
    A_{R,2}^0&=\begin{pmatrix}
        0 & 0 \\
        1 & 0 \\
        0 & 1
    \end{pmatrix}, 
    A_{R,2}^1=\begin{pmatrix}
        a & 0 \\
        0 & 1 \\
        0 & 0
    \end{pmatrix}, 
    A_{R,1}^0=\begin{pmatrix}
        0  \\
        1 
    \end{pmatrix}, 
    A_{R,1}^1=\begin{pmatrix}
        1  \\
        0 
    \end{pmatrix},\\
    A^0&=\begin{pmatrix}
        1 & 0 & 0\\
        0 & 1 & 0\\
        0 & 0 & 1
    \end{pmatrix}, 
    A^1=\begin{pmatrix}
        0 & a & 0\\
        0 & 0 & 1\\
        0 & 0 & 0
    \end{pmatrix}.
\end{align}
where $a=(C_n^2)^{-1/2}=[n(n-1)/2]^{-1/2}$ is the normalization factor. Then the SRE is ($n\ge 4$)
\begin{align}
    M_2(\ket{D^n_2}) &= -\log\left[\left(\sum_{t=0}^3E_{L,1}^{t\ \otimes 4}\right)
    \left(\sum_{t=0}^3E_{L,2}^{t\ \otimes 4}\right)
    \left(\sum_{t=0}^3E^{t\ \otimes 4}\right)^{n-4}
    \left(\sum_{t=0}^3E_{R,2}^{t\ \otimes 4}\right)
    \left(\sum_{t=0}^3E_{R,1}^{t\ \otimes 4}\right)
    \right]-n\log2.
\end{align}

Denoting 
\begin{align}
    D = \sum_{t=0}^3 (E^t)^{\otimes 4}, 
\end{align}
one can find that the order of $D$ is $3^4=6561$ thus making it quite slow for Mathematica to 
give an explicit expression of SRE. However, we can obtain the final result by using some tricks. 

Suppose $P\in\mathcal{P}_n$ is a $n$-qubit Pauli string, the following equation yields
\begin{align}
    \langle P\rangle = \bra{D^n_2}P\ket{D^n_2}=a^{2}\sum_{i=1}^{n-1}\sum_{i'=i+1}^n\sum_{j=1}^{n-1}\sum_{j'=j+1}^n
    \bra{0}^{\otimes n}\hat{\sigma}^x_{i}\hat{\sigma}^x_{i'} P \hat{\sigma}^x_{j}\hat{\sigma}^x_{j'}\ket{0}^{\otimes n},
\end{align}
and the $||\Xi(\ket{D^n_2})||$ is 
\begin{align}
    ||\Xi(\ket{D^n_2})|| = \sum_{P\in\mathcal{P}_n} \frac{\langle P\rangle^4}{2^{2n}}, \label{Eq27}
\end{align}
therefore we first conclude that 
\begin{align}
    ||\Xi(\ket{D^n_2})||= a^8 f(n)
\end{align}
where $f(n)$ is a function of $n$. 

We denote $B^t = \sqrt{2}E^t = \sum_{s,s'=0}^1\sigma^t_{s,s'}A^s\otimes \bar{A}^{s'}$ and set $a=1$. Because we have 
already known the relation of $||\Xi(\ket{D^n_2})||$ and $a$, so setting $a=1$ has no effect on $f(n)$. Now all 
the $A^s$ tensors are composed of $0$s and $1$s thus making $B^t$ tensors composed of $\pm1$ and $\pm i$ as shown below. 
\begin{align}
    B^0&=\mathbb{I}+X, B^1 = J+Y\\
    B^3&=\mathbb{I}-X, B^2=-i(J-Y)
\end{align}
where $\mathbb{I}$ is identity and 
\begin{align}
    (X)_{i,j} &= \delta_{j-i,4}-\delta_{i,3}\delta_{j,7},\quad 1\le i\le 5, 5\le j\le 9\\
    (J)_{i,j} &= \delta_{j-i,1}-\delta_{i,3}\delta_{j,4}-\delta_{i,6}\delta_{j,7}, \quad 1\le i\le 8,2\le j\le 9\\
    (Y)_{i,j} &= \delta_{j-i,3},\quad 1\le i\le 6, 4\le j\le 9.
\end{align}
We further calculate $(B^0)^{\otimes 4}+(B^3)^{\otimes 4}$ and find 
\begin{align}
    (B^0)^{\otimes 4}+(B^3)^{\otimes 4} &= (\mathbb{I}+X)^{\otimes 4}+(\mathbb{I}-X)^{\otimes 4}\\
    &= 2(\mathbb{I}^{\otimes 4}+\sum_{cyc}\mathbb{I}^{\otimes 2}X^{\otimes 2}+X^{\otimes 4})
\end{align}
and similarly
\begin{align}
    (B^1)^{\otimes 4}+(B^2)^{\otimes 4} &= (J+Y)^{\otimes 4}+(J-Y)^{\otimes 4}\\
    &= 2(J^{\otimes 4}+\sum_{cyc}J^{\otimes 2}Y^{\otimes 2}+Y^{\otimes 4})
\end{align}
where $\sum_{cyc}$ means sum over all possible permutations. 

Therefore, it can be clearly seen that the matrix $\sum_{t=0}^3 (B^t)^{\otimes 4}$ is composed of $0$s and $2$s 
because there are only $0$s and $1$s in $\mathbb{I},X,J,Y$ and they satisfy $A_{i,j}B_{i,j}=0, \forall i,j$
and thus we have
\begin{align}
    D = \sum_{t=0}^3 (E^t)^{\otimes 4} = \frac{1}{2}\left[\frac{1}{2}\sum_{t=0}^3(B^t)^{\otimes 4}\right]:=\frac{1}{2} T
\end{align}
where $T$ is only composed of $0$s and $1$s. The calculation of $||\Xi(\ket{D^n_2})||$ involves evaluating the 
matrix element of $D^n=2^{-n} T^n$, so we second conclude that 
\begin{align}
    ||\Xi(\ket{D^n_2})||= a^8 2^{-n} g(n)
\end{align}
where $g(n)$ is the combination of some of the matrix elements of $T^n$. 

For $T$, we can decomposed it into the sum of an identity $\mathbb{I}$ and a strict upper triangular matrix $U$
\begin{align}
    T = \mathbb{I}+U, 
\end{align}
and by Cayley–Hamilton theorem, $U$ is nilpotent, which means there exists a $n_m$ such that 
\begin{align}
    U^n=0, \forall n\ge n_m.
\end{align}
Therefore for $T^n$, we can expand it as 
\begin{align}
    T^n = (\mathbb{I}+U)^n = \sum_{k=0}^{\min(n,n_m-1)}C_n^k U^n = \mathbb{I}+C_n^1U+C_n^2 U^2+\cdots
\end{align}
and the matrix element of $(T^n)_{i,j}$ is 
\begin{align}
    (T^n)_{i,j} = \delta_{i,j}+C_n^1 U_{i,j}+C_n^2 (U^2)_{i,j} +\cdots
\end{align}
where all of the matrix elements of $U,U^2,\cdots$ are independent of $n$ as they are 
settled when $U$ is given. 
Now, thirdly we conclude that the matrix elements of $T^n$ must be polynomials in $n$, 
and so must $g(n)$.

Since we know $g(n)$ is a polynomial, instead of calculating $T^n$, it is easier to calculate the first few values of $g(n)$ and guess its 
expression. For example, we have the follwoing terms for $g(n)$
\begin{align}
    g(4)&=720,
    g(5)=3160,
    g(6)=9045,
    g(7)=20601, \nonumber \\
    g(8)&=40600,
    g(9)=72360, 
    g(10)=119745,\cdots
\end{align}
and if you pay enough attention, you will notice that the fourth finite difference of this sequence is constant which is 
$546$. This directly tells us $g(n)$ is a fourth-degree polynomial in $n$ and its expression can be solved by those values above. 
The final result is 
\begin{align}
    g(n) = \frac{1}{4}n(n-1)(91n^2-427n+492).
\end{align}
and thus we obtain the analytical expression of $M_2(\ket{D^n_2})$, 
\begin{align}
    M_2(\ket{D^n_2})=3\log n(n-1) -\log (91n^2-427n+492) - 2\log2, 
\end{align}
and similarly for $\ket{D^n_3}$, 
\begin{align}
    M_2(\ket{D^n_3})=3\log n(n-1)(n-2)-\log (1645n^3-18921n^2+71708n-89244)-2\log 6 . 
\end{align}
We note that one can use this method 
to give analytical expression of 2-SRE for $k\ge 3$ Dicke states.

\section{Pauli-iMPS algorithm}\label{Pauli_iMPS}
Here we present the detailed procedure of the Pauli-iMPS method. For generality, we consider an iMPS with $l$ tensors in the unit cell, i.e., 
\begin{align}
    \ket{\psi} = \sum_{\cdots, s_1,s_2, \cdots ,s_l,\cdots}\cdots A_1^{s_1} A_2^{s_2} \cdots A_l^{s_l}\cdots\ket{\cdots s_1 s_2\cdots s_l\cdots}, 
\end{align}
where the sequence of tensors $\{A_i^{s_i}\}:=(A_1^{s_1}, A_2^{s_2}, \cdots, A_l^{s_l})$ forms a single unit cell. 

\begin{enumerate}
    \item We first turn $\{A_i^{s_i}\}$ tensors into left canonical form, obtaining $\{A_{L, i}^{s_i}\}$ and bond center matrix $C$. 
    \begin{align}
        \{A_1^{s_1}, A_2^{s_2}, \cdots, A_l^{s_l}\}\to \{A_{L,1}^{s_1}, A_{L,2}^{s_2}, \cdots, A_{L,l}^{s_l}, C\}
    \end{align}
    \includegraphics[width=0.9\linewidth]{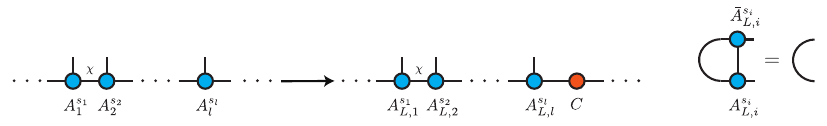}
    
    Each tensor in $\{A_{L, i}^{s_i}\}$ satisfies $\sum_{s_i}(A_{L,i}^{s_i})^{\dagger}A_{L,i}^{s_i}=\mathbb{I}$. 
    The algorithm of such transformation can be found in \cite{Vanderstraeten2019}. 

    \item We then construct $\{E_{L,i}^{t_i}\}$ tensors and denote $E_C = \bar{C}\otimes C$.
    The $E$ tensors are defined as (left side of the following picture)
    \begin{align}
        E_{L,i}^{s_i} = \sum_{s_i,s_i'=0}^1\frac{\sigma^{t_i}_{s_i,s_i'}}{\sqrt{2}}(\bar{A}_{L,i}^{s_i'}\otimes A_{L,i}^{s_i}). 
    \end{align}
    \includegraphics[width=0.9\linewidth]{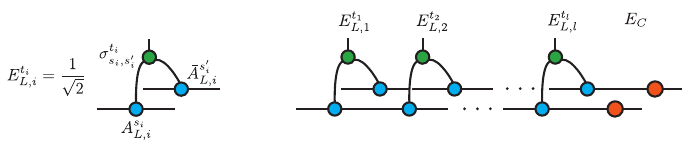}

    Now we have the tensors of Pauli iMPS $\{E_{L,i}^{s_i}\}, E_C$ without any truncation (right side of the above picture). 

    \item To reduce the bond dimension of Pauli iMPS, which is $\chi^2$ now, we perform SVD on $E_C$ 
    and truncate small singular values smaller than cutoff $\epsilon_1$ (or retain bond dimension up to $d$),
    obtaining
    \begin{align}
        E_C = \bar{C}\otimes C = 
        U_{\chi^2\times \chi^2} S_{\chi^2\times \chi^2} V^{\dagger}_{\chi^2\times \chi^2}\approx 
        \tilde{U}_{\chi^2\times d} \tilde{S}_{d\times d} \tilde{V}^{\dagger}_{d\times \chi^2} = 
        \tilde{U}_{\chi^2\times d} (\tilde{E}_C)_{d\times d} V^{\dagger}_{d\times \chi^2}, 
    \end{align}
    \noindent\makebox[0.9\linewidth][c]{%
    \includegraphics[width=0.7\linewidth]{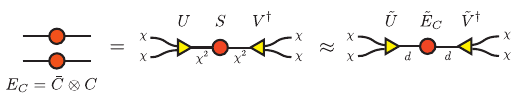}
    }

    where we denote $\tilde{E}_C:=\tilde{S}$. 

    \item We employ the unitary operator $\tilde{U}$ to reduce the bond dimension of $\{E_{L, i}^{t_i}\}$ tensors through the following transformation: 
    \begin{align}
        E_{L, i}^{t_i} \to \tilde{E}_{L, i}^{t_i} = \tilde{U}^{\dagger} E_{L, i}^{t_i} \tilde{U}. 
    \end{align}
    \includegraphics[width=0.9\linewidth]{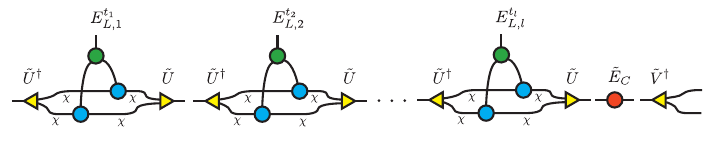}

    In the absence of truncation, the transformation is purely a gauge transformation, as it corresponds to inserting an identity between the $E$ tensors and thus preserves all physical quantities of the Pauli iMPS. 

    \item Similar to step 2, we use the truncated $E$ tensors $\{\tilde{E}_{L,i}^{t_i}\}$ to 
    construct $G$ tensors $\{G_{L, i}^{t_i}\}$. The $G_{L, i}^{t_i}$ is defined as (left side of the picture)
    \begin{align}
        G_{L, i}^{t_i} = \sum_{t_i',t_i''}\delta^{t_i}_{t_i',t_i''}
        (\tilde{E}_{L, i}^{t_i'}\otimes \tilde{E}_{L, i}^{t_i''}).
    \end{align}
    \includegraphics[width=0.9\linewidth]{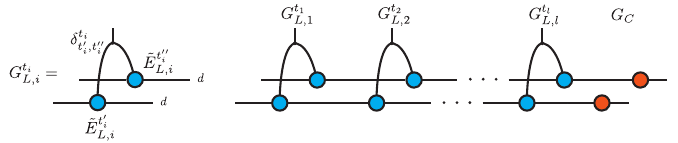}

    And the $G_C$ tensor is defined as $G_C = \tilde{E}_C\otimes \tilde{E}_C$. Here $G_C$ tensor is a diagonal matrix because $\tilde{E}_C$ is diagonal. 

    \item Similar to step 3, we perform SVD on $G_C$ and truncate small singular values smaller than cutoff $\epsilon_2$ (or retain bond dimension up to $m$), obtaining
    \begin{align}
        G_C = \tilde{E}_C\otimes \tilde{E}_C = U_{d^2\times d^2}S_{d^2\times d^2}V^{\dagger}_{d^2\times d^2}
        \approx U_{d^2\times m}S_{m\times m}V^{\dagger}_{m\times d^2} = 
        U_{d^2\times m}(\tilde{G}_C)_{m\times m}V^{\dagger}_{m\times d^2}.
    \end{align}
    \noindent\makebox[0.9\linewidth][c]{%
    \includegraphics[width=0.7\linewidth]{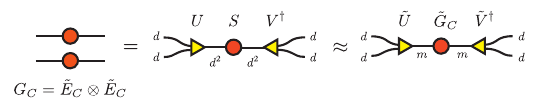}
    }
    
    Still, we denote $\tilde{G}_C:= S_{m\times m}$. 

    \item Similar to step 4, we employ the unitary operator $\tilde{U}$ to reduce the bond dimension of 
    $\{G_{L, i}^{t_i}\}$ tensors through
    \begin{align}
        G_{L, i}^{t_i} \to \tilde{G}_{L, i}^{t_i} = \tilde{U}^{\dagger} G_{L, i}^{t_i} \tilde{U}. 
    \end{align}
    \includegraphics[width=0.9\linewidth]{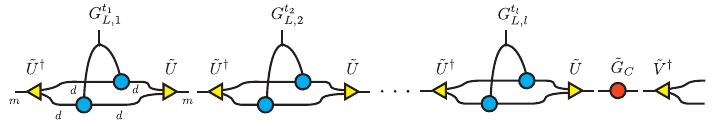}

    \item Now we obtain truncated $G$ tensors $\{\tilde{G}_{L, i}^{t_i}\}$. We use these tensors to 
    construct $D_L$ tensor, defined as 
    \begin{align}
        D_L = \left(\sum_{t_1}\tilde{G}_{L, 1}^{t_1}\otimes\tilde{G}_{L, 1}^{t_1}\right) 
        \left(\sum_{t_2}\tilde{G}_{L, 2}^{t_2}\otimes\tilde{G}_{L, 2}^{t_2}\right) \cdots 
        \left(\sum_{t_l}\tilde{G}_{L, l}^{t_l}\otimes\tilde{G}_{L, l}^{t_l}\right) .
    \end{align}
    \noindent\makebox[0.9\linewidth][c]{%
    \includegraphics[width=0.45\linewidth]{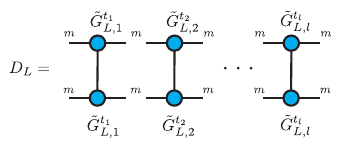}
    }

    Our goal is to compute the dominant eigenvalue of $D_L$, denoted by $\lambda_0(D_L)$. 
    When $m$ is moderately small, this can be done straightforwardly. However, to achieve higher accuracy, 
    $m$ may need to reach $O(10^4)$. In this regime, explicitly constructing $D_L$ becomes computationally prohibitive. 
    Instead, we determine $\lambda_0(D_L)$ using Krylov subspace method by implementing only the action of $D_L$ on vectors, i.e., by treating $D_L$ as an abstract linear map rather than forming it as a dense matrix. This approach significantly reduces the computational cost. 

    \item After obtaining $\lambda_0(D_L)$, the SRE density of the iMPS $\ket{\psi}$ with $l$ tensors in single 
    unit cell can be calculated, 
    \begin{align}
        m_2(\ket{\psi}) = -\frac{1}{l}\log|\lambda_0(D_L)| - \log 2.  \label{multi-site_SRE_density}
    \end{align}
\end{enumerate}

\section{Bond-DMRG algorithm}\label{MPO_construct}
\subsection{Construction of $D$ MPO}
We directly construct the MPO form of $D$. Here we consider there is only one tensor $A^s$ in the unit cell. 
The $A^s$ tensor doesn't need to be in the left canonical form. 
We now define new transfer matrices, which is basically $E^t$ but with no $1/\sqrt{2}$ factor, i.e., 
\begin{align}
    B^t = \sqrt{2}E^t = \sum_{s,s'=0}^{1}\sigma^t_{s,s'}A^s\otimes\bar{A}^{s'}
\end{align}
and the four $B$ matrices can be clearly written, 
\begin{align}
    B^0 &= A^0\otimes\bar{A}^0+A^1\otimes\bar{A}^1\\
    B^1 &= A^0\otimes\bar{A}^1+A^1\otimes\bar{A}^0\\
    B^2 &= -iA^0\otimes\bar{A}^1+iA^1\otimes\bar{A}^0\\
    B^3 &= A^0\otimes\bar{A}^0-A^1\otimes\bar{A}^1.
\end{align}

First define $F^{t}$ as 
\begin{align}
    F^{0} = A^0\otimes\bar{A}^0, F^{1} = A^0\otimes\bar{A}^1\\
    F^{2} = A^1\otimes\bar{A}^0, F^{3} = A^1\otimes\bar{A}^1
\end{align}
and now we put $B^0$ and $B^3$ together, 
\begin{align}
    (B^0)^{\otimes 4} + (B^3)^{\otimes 4} &= (F^0+F^3)^{\otimes 4} + (F^0-F^3)^{\otimes 4}\\
    &= 2\left[(F^0)^{\otimes4} + (F^3)^{\otimes4} + 
    F^0\otimes F^0\otimes F^3\otimes F^3 \right.\notag\\
    &+F^0\otimes F^3\otimes F^0\otimes F^3 + 
    F^0\otimes F^3\otimes F^3\otimes F^0 \notag\\
    &\left.+F^3\otimes F^0\otimes F^0\otimes F^3 + 
    F^3\otimes F^0\otimes F^3\otimes F^0 + 
    F^3\otimes F^3\otimes F^0\otimes F^0 \right]
\end{align}
and this term can be written in the MPO form with bond dimension $2$, 
\begin{align}
    \frac12\left[(B^0)^{\otimes 4} + (B^3)^{\otimes 4}\right] = \begin{bmatrix}
        F^3 & F^0
    \end{bmatrix}
    \begin{bmatrix}
        F^0 & F^3\\
        F^3 & F^0
    \end{bmatrix}
    \begin{bmatrix}
        F^0 & F^3\\
        F^3 & F^0
    \end{bmatrix}
    \begin{bmatrix}
        F^3 \\ F^0
    \end{bmatrix}
\end{align}
and similarly for $(B^1)^{\otimes 4} + (B^2)^{\otimes 4}$ we have
\begin{align}
    \frac12\left[(B^1)^{\otimes 4} + (B^2)^{\otimes 4}\right] &= (F^1+F^2)^{\otimes 4} + (F^1-F^2)^{\otimes 4}\\
    &= \begin{bmatrix}
        F^2 & F^1
    \end{bmatrix}
    \begin{bmatrix}
        F^1 & F^2\\
        F^2 & F^1
    \end{bmatrix}
    \begin{bmatrix}
        F^1 & F^2\\
        F^2 & F^1
    \end{bmatrix}
    \begin{bmatrix}
        F^2 \\ F^1
    \end{bmatrix}.
\end{align}

Then the MPO representation of $4D = \sum_{t=0}^3(B^t)^{\otimes 4}$ is 
\begin{align}
    D = 
    \frac14\sum_{t=0}^3(B^t)^{\otimes 4} = \frac12
    \begin{bmatrix}
        F^3 & F^0 & F^2 & F^1
    \end{bmatrix}
    \begin{bmatrix}
        F^0 & F^3 & 0 & 0\\
        F^3 & F^0 & 0 & 0\\
        0 & 0 & F^1 & F^2\\
        0 & 0 & F^2 & F^1
    \end{bmatrix}
    \begin{bmatrix}
        F^0 & F^3 & 0 & 0\\
        F^3 & F^0 & 0 & 0\\
        0 & 0 & F^1 & F^2\\
        0 & 0 & F^2 & F^1
    \end{bmatrix}
    \begin{bmatrix}
        F^3 \\ F^0 \\ F^2 \\ F^1
    \end{bmatrix}
\end{align}
and further we can write the MPO of $D$ in terms of $A$ tensor with bond dimension no more than 8, 
\begin{align}
    D = \frac{1}{2}\begin{bmatrix}
        1 & 0 & 1 & 0
    \end{bmatrix}
    \begin{bmatrix}
        \bar{1} & \cdot & \cdot & \cdot \\
        \cdot & \bar{0} & \cdot & \cdot \\
        \cdot & \cdot & \bar{0} & \cdot \\
        \cdot & \cdot & \cdot & \bar{1} \\
    \end{bmatrix}
    \begin{bmatrix}
        0 & 1 & \cdot & \cdot & \cdot & \cdot & \cdot & \cdot \\
        \cdot & \cdot & 1 & 0 & \cdot & \cdot & \cdot & \cdot \\
        \cdot & \cdot & \cdot & \cdot & 0 & 1 & \cdot & \cdot \\
        \cdot & \cdot & \cdot & \cdot & \cdot & \cdot & 1 & 0 \\
    \end{bmatrix}
    \begin{bmatrix}
        \bar{0} & \cdot & \cdot & \cdot \\
        \cdot & \bar{1} & \cdot & \cdot \\
        \bar{1} & \cdot & \cdot & \cdot \\
        \cdot & \bar{0} & \cdot & \cdot \\
        \cdot & \cdot & \bar{1} & \cdot \\
        \cdot & \cdot & \cdot & \bar{0} \\
        \cdot & \cdot & \bar{0} & \cdot \\
        \cdot & \cdot & \cdot & \bar{1} \\
    \end{bmatrix}
    \begin{bmatrix}
        0 & 1 & \cdot & \cdot & \cdot & \cdot & \cdot & \cdot \\
        \cdot & \cdot & 1 & 0 & \cdot & \cdot & \cdot & \cdot \\
        \cdot & \cdot & \cdot & \cdot & 0 & 1 & \cdot & \cdot \\
        \cdot & \cdot & \cdot & \cdot & \cdot & \cdot & 1 & 0 \\
    \end{bmatrix}
    \begin{bmatrix}
        \bar{0} & \cdot & \cdot & \cdot \\
        \cdot & \bar{1} & \cdot & \cdot \\
        \bar{1} & \cdot & \cdot & \cdot \\
        \cdot & \bar{0} & \cdot & \cdot \\
        \cdot & \cdot & \bar{1} & \cdot \\
        \cdot & \cdot & \cdot & \bar{0} \\
        \cdot & \cdot & \bar{0} & \cdot \\
        \cdot & \cdot & \cdot & \bar{1} \\
    \end{bmatrix}
    \begin{bmatrix}
        1 & \cdot & \cdot & \cdot \\
        \cdot & 0 & \cdot & \cdot \\
        \cdot & \cdot & 1 & \cdot \\
        \cdot & \cdot & \cdot & 0 \\
    \end{bmatrix}
    \begin{bmatrix}
        \bar{1} \\ \bar{0} \\ \bar{0} \\ \bar{1}
    \end{bmatrix}
\end{align}
where we use $0, 1, \bar{0}, \bar{1}$ to represent $A^0, A^1, \bar{A}^0$ and $\bar{A}^1$, and $\cdot$ for zero matrix, 
respectively. 

\begin{figure}
    \centering
    \includegraphics[width=0.4\linewidth]{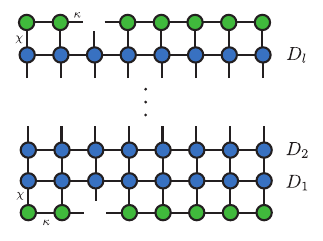}
    \caption{The basic idea of bond-DMRG algorithm is to construct the MPO form of matrix $D$ and perform standard DMRG to obtain the dominant eigenvalue. For $l$-site unit cell iMPS, 
    the MPO has $l$ layers. 
    The bond dimension of matrix $D$'s MPO is no more than 8.}
    \label{bDMRG_alg}
\end{figure}

\subsection{The multi-site unit cell case}
For an $l$-site unit-cell iMPS, $\{A_i^{s_i}\}_{i=1,\cdots,l}$, the corresponding $D$ MPO, denoted by $\{D_i\}$, can be constructed following the procedure described above. We then apply DMRG to the resulting multi-layer MPO, as illustrated in Fig.~\ref{bDMRG_alg}. The SRE density is subsequently obtained according to Eq.~\eqref{multi-site_SRE_density}. We note that the MPO is not hermitian in general cases, so one should use Arnoldi method as 
the local eigensolver in the DMRG procedure. A better way may be to employ biorthonormal-block
DMRG \cite{Zhong2025}. However, we found standard DMRG is enough for our purpose in most case.

\section{Performance of Pauli-iMPS and bond-DMRG methods}
\subsection{Pauli-iMPS method}

We first study how truncation parameters $d, m$ affect the result of Pauli-iMPS method. 
Therefore, we fix the bond dimension $\chi$ of $A^s$ and 
solve for ground state of 1D Ising model at $h_x = 0.9, 1$ and $1.1$ via VUMPS. 
For $\chi=4$ and $5$, the SRE density can be accurately obtained by setting the truncation parameters to 
$(d, m)=(\chi^2, \chi^4)$. 
We denote by $m_2(\chi,d,m)$ the SRE density calculated with truncation parameters $\chi,d$ and $m$.
Under a fixed bond dimension $\chi$, the reference (accurate) value is therefore given by $m_2(\chi,\chi^2,\chi^4)$. 
We also introduce the notation $m_2(a, b):=m_2(\chi,\chi^a,\chi^{ab})$, 
where $a=\log_{\chi}d$ and $b = \log_d m$. In the calculation, we set $a, b\in[1, 2]$. 

\begin{figure}
    \centering
    \includegraphics[width=1.0\linewidth]{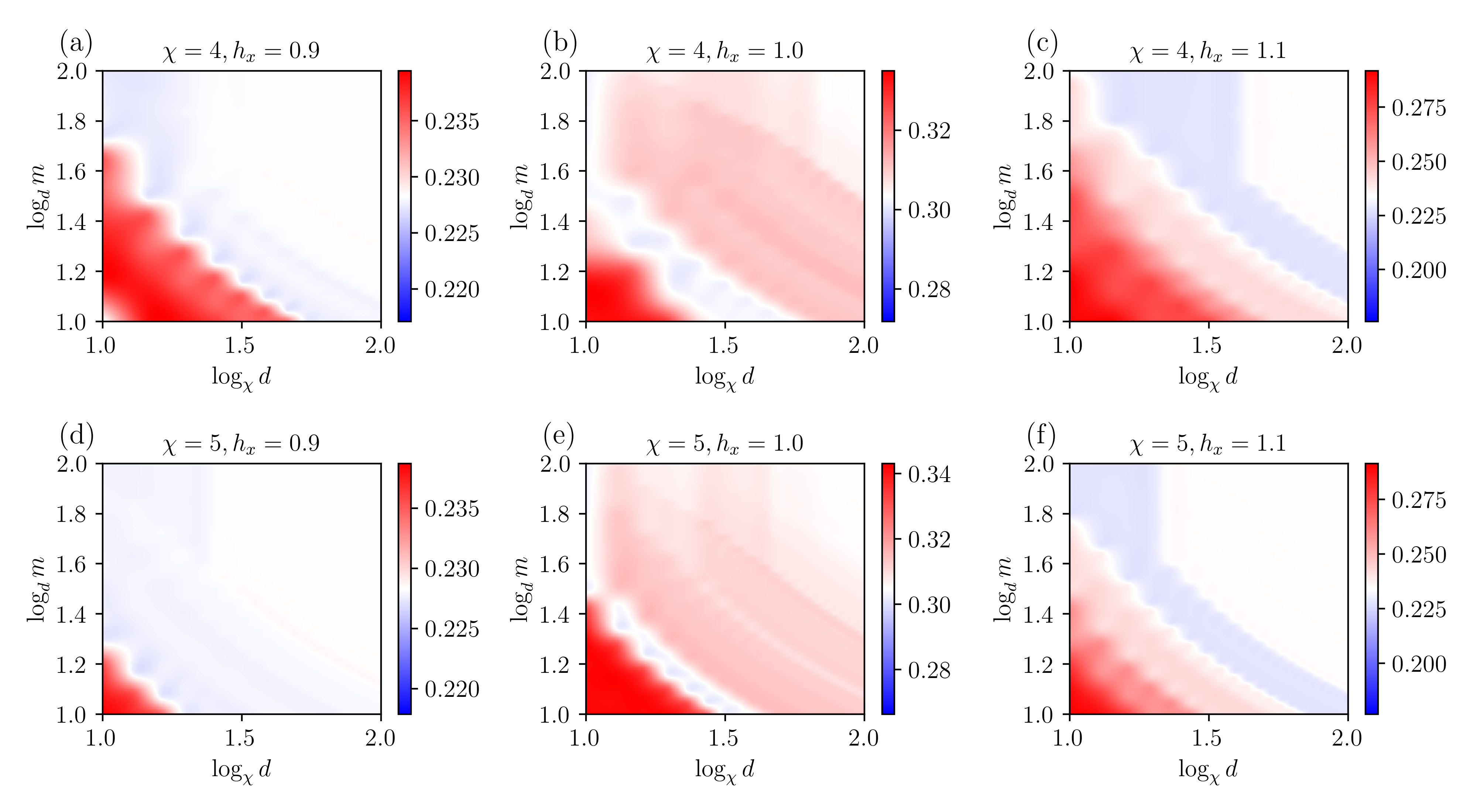}
    \caption{SRE density $m_2$ in the $\log_{\chi} d$-$\log_d m$ plane with different $\chi$ and $h_x$. 
    The color bar is centered at the reference value $m_2(2,2)$ (white). Red (blue) indicates values larger (smaller) than the reference.}
    \label{pimps1}
\end{figure}

It can be observed that small values of $d$ and $m$ tend to yield results that are larger than the reference value in most cases, as shown in Fig.~\ref{pimps1}. This effect becomes more pronounced near the critical point and in the paramagnetic phase. 
At the critical point, the red region remains essentially unchanged when increasing $\chi$ from $4$ to $5$. 
In the paramagnetic phase ($h_x = 1.1$), the red region shrinks slightly, but the reduction is much less significant than in the ferromagnetic phase ($h_x=0.9$). 
These observations suggest that achieving an accurate evaluation of the SRE density in the disordered (symmetry-unbroken) phase requires greater computational effort.

\begin{figure}
    \centering
    \includegraphics[width=0.75\linewidth]{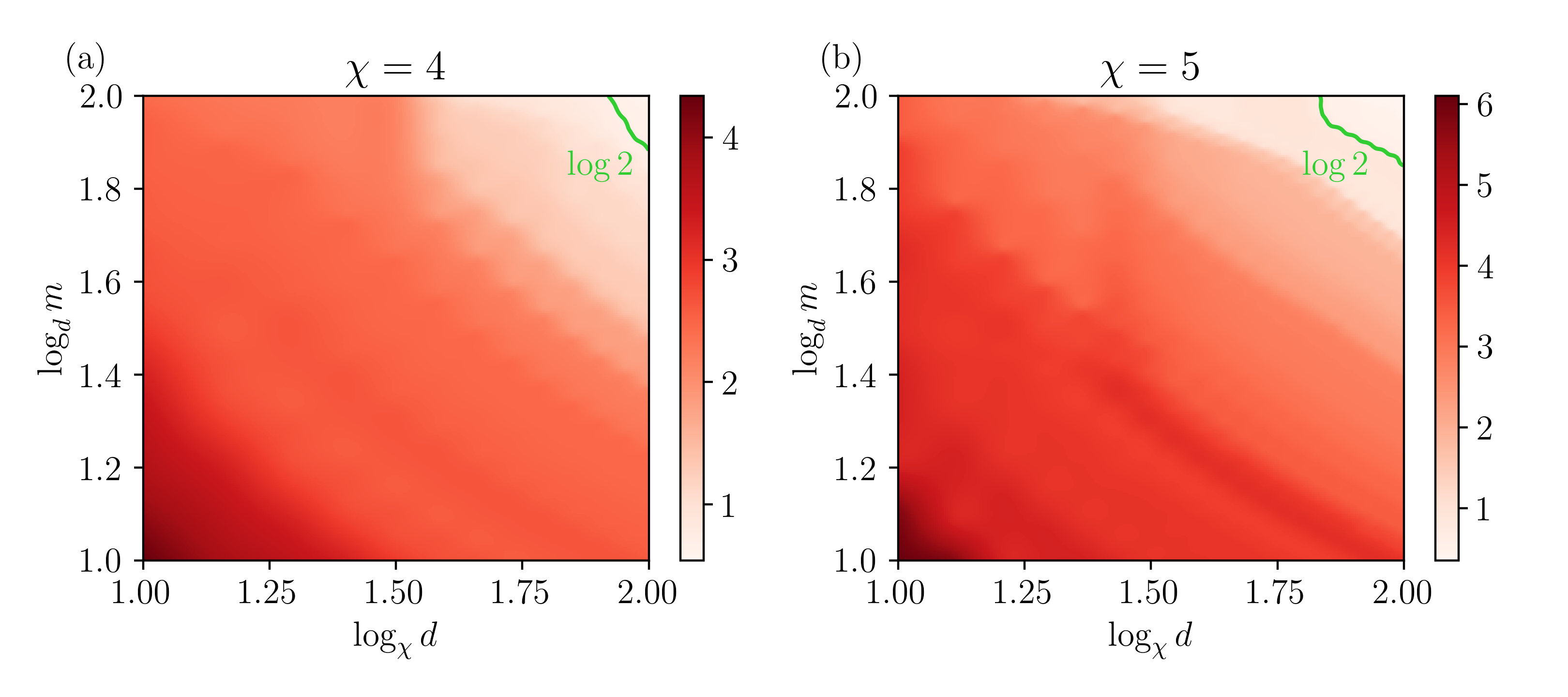}
    \caption{SRE density $m_2$ in the $\log_{\chi} d$-$\log_d m$ plane with different $\chi$ for two randomly generated iMPS (constructed following the method described in the paper). In the color bar, white denotes the accurate reference value, while deeper red corresponds to larger deviations. The green line represents the theoretical upper bound $\log2$. 
    }
    \label{pimps2}
\end{figure}

For random iMPS, this effect cannot be neglected. Moreover, in such cases, any truncation may lead to significant deviations in the final results. 
It is evident that when the truncation becomes moderately strong ($\log_{\chi}d+\log_dm \lesssim 3.9$), the results may become unphysical, exceeding the theoretical upper bound $\log 2$. This indicates that the Pauli-MPS method performs poorly for random iMPS. 

\begin{figure}
    \centering
    \includegraphics[width=0.9\linewidth]{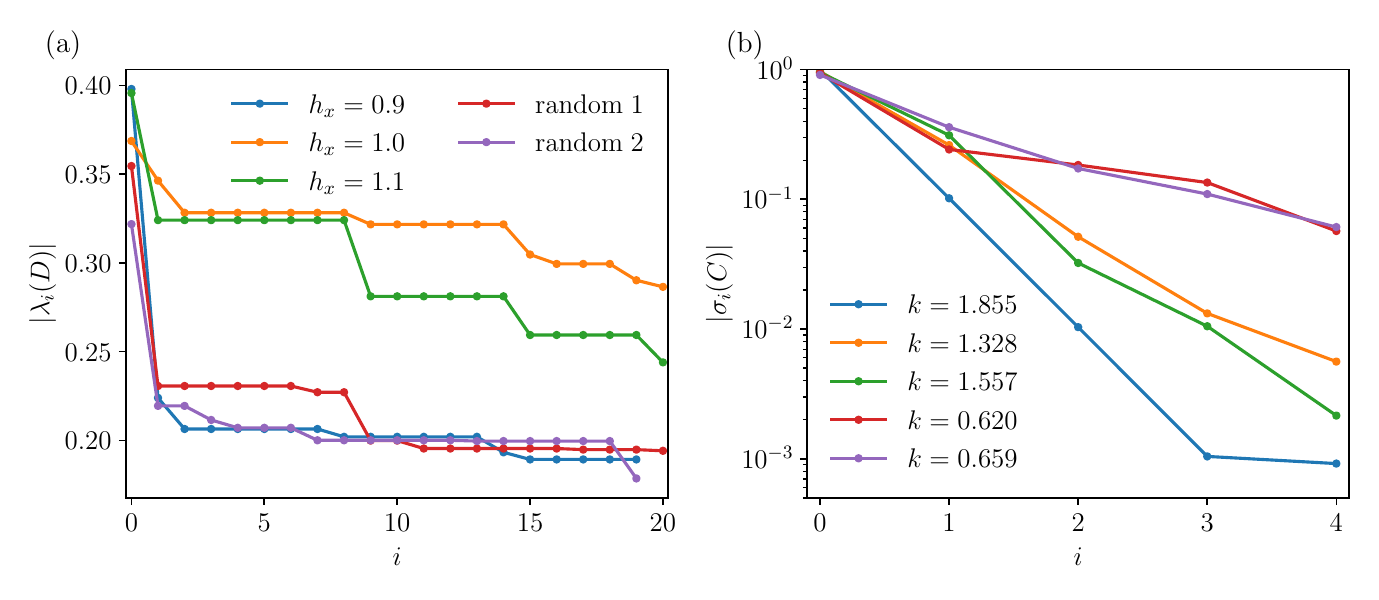}
    \caption{
    (a) Several of the largest eigenvalues of the transfer matrix $D$, denoted by $\lambda_i(D)$, are shown. Random 1 corresponds to the $\chi = 5$ iMPS presented in Fig.~\ref{pimps2}.
    (b) The singular values of the bond-center matrix $C$ for different iMPS are displayed. The color scheme in panel (b) follows that in panel (a), where each color represents the same iMPS. 
    Here we fix $\chi = 5$ and perform no truncation in panels (a) and (b). The decay rate $k$ is obtained from a fit in the natural exponential form $\ln |\sigma_x(C)| \sim -k x$; that is, the logarithm is taken with base $e$, not base 10.
    }
    \label{spectrums}
\end{figure}

To understand this poor performance, we compute several of the largest eigenvalues of the transfer matrix $D$, as well as the singular values of the bond-center matrix $C$, for different iMPS. The results are shown in Fig.~\ref{spectrums}.
From Fig.~\ref{spectrums} (a), we observe that the spectrum of $D$ for the Ising model in the ferromagnetic phase is similar to that of the random iMPS, while the spectra in the critical and paramagnetic phases exhibit similar structures. This suggests that the spectrum of $D$ may be closely related to the symmetry properties of the iMPS, since the states at $h_x = 1.0$ and $1.1$ preserve the underlying symmetry.
Figure~\ref{spectrums} (b) shows the singular values of $C$ for different iMPS. Assuming an approximate exponential decay of the form $|\sigma_x(C)| \sim e^{-k x}$, we extract the decay rate $k$ from fitting, as indicated in Fig.~\ref{spectrums} (b). We find that the random iMPS exhibits the smallest decay rate, whereas the iMPS in the ferromagnetic phase shows the largest decay rate. The cases at $h_x = 1.0$ and $1.1$ lie between these two extremes.
Importantly, we observe that iMPS with a larger decay rate allow for a more reliable extraction of the SRE density using the Pauli-iMPS method, even in the presence of substantial truncation, for example, the state at $h_x = 0.9$. In contrast, for random iMPS, even small truncation errors can lead to significant deviations in the SRE density. This is because the slow decay of singular values implies that truncation discards a considerable amount of relevant information. 

Finally, we analyze the computational complexity of the Pauli-iMPS method. The most time-consuming step is the computation of the dominant eigenvalue of the matrix $D$. To accomplish this, we employ Krylov subspace methods, whose core operation is repeated matrix-vector multiplication. In our implementation, each matrix-vector multiplication corresponds to contracting two $\tilde{G}_L^t$ tensors with a trial matrix.
As a result, the computational cost of the Pauli-iMPS method scales as $O(m^3)$. If no truncation is performed, the effective bond dimension satisfies $m \sim \chi^4$, leading to a time complexity of $O(\chi^{12})$.
Empirically, as observed in Fig.~\ref{pimps1}, obtaining reliable results for one-dimensional spin models requires $\log_{\chi} d + \log_d m \ge 2\sqrt{\log_{\chi} m} \gtrsim 3.5$, which implies $m \gtrsim \chi^3$. Under this condition, the overall computational complexity scales as $O(\chi^9)$.
However, when computing the SRE density for random iMPS, truncation generally leads to unreliable results. In this case, one must effectively avoid truncation, and the computational cost remains $O(\chi^{12})$.

\subsection{bond-DMRG method}
The only truncation parameter in bond-DMRG is the bond dimension $\kappa$ of the MPS. So, we study 
how bond dimension $\kappa$ affect the result. We take the $m_2^{\mathrm{ref}} = m_2(2,2)$ 
from Pauli-iMPS method as the reference value and define the relative error of the bond-DMRG result as 
\begin{align}
    \epsilon = \frac{|m_2^{\mathrm{bDMRG}}-m_2^{\mathrm{ref}}|}{m_2^{\mathrm{ref}}}
\end{align}
where $m_2^{\mathrm{bDMRG}}$ is the SRE density from bond-DMRG method. 

\begin{figure}
    \centering
    \includegraphics[width=0.97\linewidth]{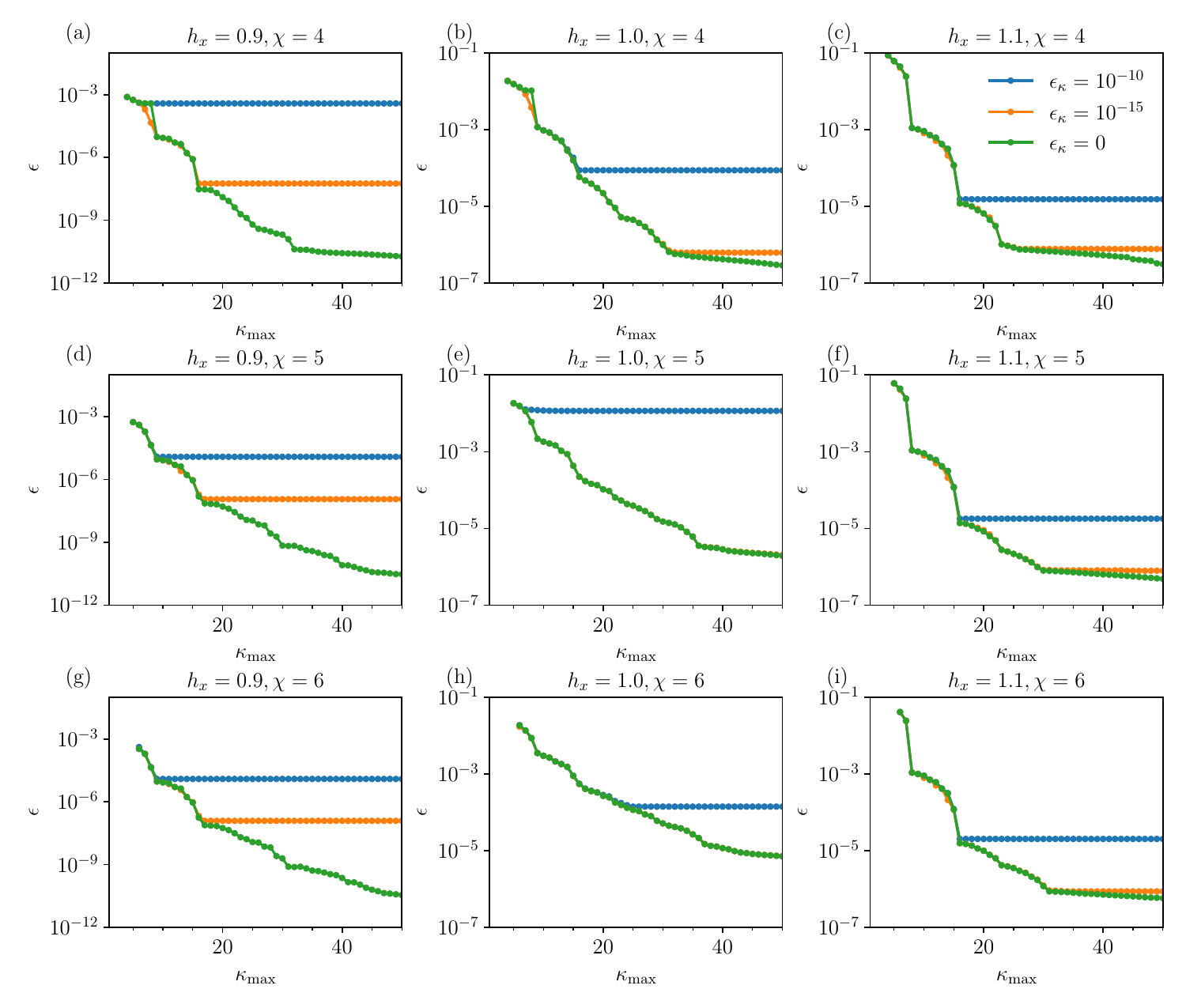}
    \caption{Relative error $\epsilon$ of bond-DMRG results obtained with various maximum bond dimensions $\kappa_{\mathrm{max}}$. Different colors correspond to different singular-value truncation cutoffs,  where singular values smaller than $\epsilon_{\kappa}$ are discarded. 
    }
    \label{bdmrg1}
\end{figure} 

The results for different values of $h_x$ in the Ising model are shown in Fig.~\ref{bdmrg1}. It is evident that the bond-DMRG approach yields significantly higher accuracy compared to the Pauli-iMPS method with truncation. Even when the bond dimension $\kappa$ is kept relatively small, for instance $\kappa \sim \chi^2$, the SRE density remains highly accurate, with a relative error on the order of $10^{-5}$.
Moreover, one observes that achieving the same level of accuracy at $h_x = 1.0$ requires substantially more computational resources than at $h_x = 0.9$ or $1.1$. This reflects the general fact that computing the SRE density for a critical state is more challenging than for a gapped state.

\begin{figure}
    \centering
    \includegraphics[width=1\linewidth]{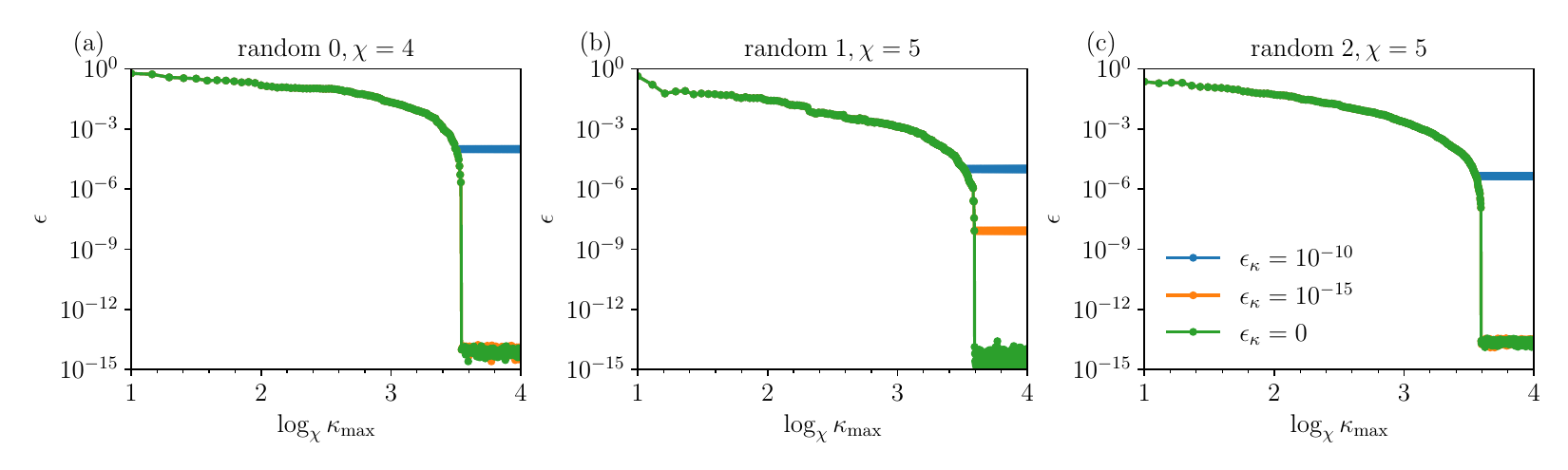}
    \caption{Relative error $\epsilon$ of the bond-DMRG results for three random iMPS. Panels (a) and (b) correspond to the same iMPS as in Fig.~\ref{pimps2} (a) and (b), respectively, while panel (c) corresponds to the ``random 2'' state in Fig.~\ref{spectrums}. Colors represent different singular-value truncation cutoffs $\epsilon_{\kappa}$, below which singular values are discarded.}
    \label{bdmrg2}
\end{figure}

However, for random iMPS, substantially more computational resources are required to obtain accurate results. As shown in Fig.~\ref{bdmrg2}, achieving high precision demands that the bond dimension scales approximately as $\kappa \sim O(\chi^{3.6})$, which quickly becomes computationally prohibitive for large $\chi$. Even for a moderate accuracy, say a relative error of order $O(10^{-3})$, one still requires $\kappa \sim O(\chi^{3})$. Therefore, determining the accurate SRE density for random iMPS is considerably more challenging than for ground states of one-dimensional spin models.

Finally, we analyze the computational complexity of the bond-DMRG method.
Treating $\chi$ as the local physical dimension and $\kappa$ as the MPS bond dimension, the time complexity scales as $O(\chi \kappa^3)$ for the one-site DMRG algorithm and $O(\chi^3 \kappa^3)$ for the two-site DMRG algorithm.
For ground states of one-dimensional spin models, a bond dimension $\kappa \sim O(\chi^2)$ is typically sufficient. In this case, the overall time complexity becomes $O(\chi^7)$ for the one-site algorithm and $O(\chi^9)$ for the two-site algorithm. Therefore, when using the one-site update scheme, bond-DMRG is more efficient than the Pauli-iMPS approach. With the two-site algorithm, the time complexity is comparable, while bond-DMRG remains more favorable in terms of memory consumption.
For random iMPS, however, a significantly larger bond dimension is required, namely $\kappa \sim O(\chi^{3\text{–}3.6})$. This leads to an overall time complexity of $O(\chi^{10\text{–}11.8})$ for the one-site algorithm and $O(\chi^{12\text{–}13.8})$ for the two-site algorithm. In this regime, the two-site algorithm becomes more expensive than the Pauli-iMPS method. The reason is that the two-site update involves a full SVD, whose computational cost scales as $O(n^3)$ for a general $n \times n$ matrix.
More efficient SVD techniques, such as randomized SVD, could in principle reduce the scaling to approximately $O(\chi^{11\text{–}12.8})$. Although the two-site algorithm may still exceed the $O(\chi^{12})$ scaling of the untruncated Pauli-iMPS method, it is worth emphasizing that bond-DMRG has a lower memory requirement, which can be advantageous in practical large-scale computations.

\subsection{Comparison between two methods}
The objective of both approaches is to obtain the dominant eigenvalue of the matrix $D$. The main challenge arises from the extremely large dimension of $D$, which scales as $\chi^8$. Both methods attempt to reduce the computational cost by compressing $D$ while minimizing the impact on its dominant eigenvalue.
The Pauli-iMPS method constructs unitary transformations based on the bond-center matrix $C$, which effectively act as projectors and reduce the dimension of $D$. However, from a conceptual standpoint, this strategy is not necessarily optimal, since the dominant eigenvalue of $D$ is not directly determined by the structure of the $C$ matrix. In principle, one could construct a more tailored compression scheme that preserves the dominant eigenvalue more efficiently under the same truncation level. Nevertheless, the current Pauli-iMPS approach is sufficiently effective for iMPS with relatively small bond dimensions.
In contrast, the bond-DMRG method exploits the variational structure of DMRG and represents the dominant eigenvector of $D$ as an MPS. Instead of explicitly constructing the full matrix $D$, one only needs to solve for the dominant eigenvalue of local effective operators, with the computational cost controlled by the MPS bond dimension. This makes the method particularly suitable for iMPS with larger bond dimensions, where extrapolation techniques can be employed when the required bond dimension exceeds available computational resources. Moreover, bond-DMRG can already achieve reasonably high precision even with relatively small bond dimensions.
In summary, the Pauli-iMPS method is more appropriate for iMPS with small bond dimensions, while the bond-DMRG method is better suited for systems with larger bond dimensions.

\section{Asymptotic behaviour of two-site mutual SRE in the injective TIMPS}\label{appendix_mSRE}
The stabilizer R\'enyi entropy $\tilde{M}_2(\rho)$ can be extended to mixed states, 
defined as 
\begin{align}
    \tilde{M}_2(\rho) := M_2(\rho)-S_2(\rho)
\end{align}
with $S_2(\rho)$ the 2-R\'enyi entropy of $\rho$ and
\begin{align}
    M_2(\rho) = -\log \mathrm{tr}(Q_n\rho^{\otimes4})-\log d.
\end{align}
where $Q_n = d^{-2}\sum_{P\in\mathcal{P}_n}P^{\otimes 4}$.

The long-range non-stabilizerness can thus be quantified by 
\begin{align}
    L(\rho_{AB}) = \tilde{M}_2(\rho_{AB}) - \tilde{M}_2(\rho_A) - \tilde{M}_2(\rho_B).
\end{align}
which consists of two parts
\begin{align}
    L(\rho_{AB}) &= L_{M}(\rho_{AB}) + L_{S}(\rho_{AB}), \\
    L_{M}(\rho_{AB}) &:= M_2(\rho_{AB}) - M_2(\rho_{A}) - M_2(\rho_{B}), \\
    L_{S}(\rho_{AB}) &:= S_2(\rho_{A}) + S_2(\rho_{B}) - S_2(\rho_{AB}).
\end{align}

We first calculate the R\'enyi entropy part. 
Suppose there are $r$ sites between $A$ and $B$(and the site of $A$ and $B$ are denoted as $s_1,s_2$, and we have $s_2-s_1-1 = r$), 
then the density matrix of joint system $AB$ is
\begin{align}
    \rho_{AB} &= (l|
    (A^{s_1}\otimes\bar{A}^{s_1'}) (E_{A}^{\bar{A}})^r (A^{s_2}\otimes\bar{A}^{s_2'})
    |r), 
\end{align}
using the eigendecomposition of transfer matrix $E_{A}^{\bar{A}}$, 
\begin{align}
    E_{A}^{\bar{A}} = |r)(l| + \sum_{i}\lambda_i|\lambda_i)(\lambda_i|
\end{align}
we obtain
\begin{align}
    \rho_{AB} = (l|(A^{s_1}\otimes\bar{A}^{s_1'})|r) \otimes (l|(A^{s_2}\otimes\bar{A}^{s_2'})|r)
    +\sum_i\lambda_i^r (l|(A^{s_1}\otimes\bar{A}^{s_1'})|\lambda_i)\otimes (\lambda_i|(A^{s_2}\otimes\bar{A}^{s_2'})|r)
\end{align}
denoting
\begin{align}
    \rho_l(\lambda_i) = (l|(A^{s_1}\otimes\bar{A}^{s_1'})|\lambda_i)\\
    \rho_r(\lambda_i) = (\lambda_i|(A^{s_2}\otimes\bar{A}^{s_2'})|r)
\end{align}
the above equation is simplified to 
\begin{align}
    \rho_{AB} = \rho_A\otimes\rho_B + \sum_i\lambda_i^r\rho_l(\lambda_i)\otimes\rho_r(\lambda_i). \label{densitymatrix}
\end{align}

Now we try to calculate the R\'enyi entropy of $\rho_{AB}$, first we square $\rho_{AB}$
\begin{align}
    \rho_{AB}^2 &= \rho_A^2\otimes\rho_B^2+\left(
        \sum_i\lambda_i^r \rho_A\rho_l(\lambda_i)\otimes\rho_B\rho_r(\lambda_i)
    \right) + 
    \left(
        \sum_i\lambda_i^r \rho_l(\lambda_i)\rho_A\otimes\rho_r(\lambda_i)\rho_B
    \right)\notag \\
    & + \left(
        \sum_{i, j}\lambda_i^r\lambda_j^r \rho_l(\lambda_i)\rho_l(\lambda_j)\otimes\rho_r(\lambda_i)\rho_r(\lambda_j)
    \right)
\end{align}
then trace it, 
\begin{align}
    \mathrm{tr}(\rho_{AB}^2) = \tr(\rho_A^2)\tr(\rho_B^2) + 2\sum_i\lambda_i^r\tr[\rho_A\rho_l(\lambda_i)]\tr[\rho_B\rho_r(\lambda_i)]
    +\sum_{i, j}\lambda_i^r\lambda_j^r\tr[\rho_l(\lambda_i)\rho_l(\lambda_j)]\tr[\rho_r(\lambda_i)\rho_r(\lambda_j)]
\end{align}
we now obatin the R\'enyi entropy of $\rho_{AB}$
\begin{align}
    S_2(\rho_{AB}) = -\log\left[\tr(\rho_{AB}^2)\right]
\end{align}
and the expression for $L_S(\rho_{AB})$ is 
\begin{align}
    L_S(\rho_{AB}) &= - 2\log[\tr(\rho_A^2)] + \log\left[\tr(\rho_{AB}^2)\right] \\
    &= \log \left[
        1+2\sum_i\lambda_i^r\frac{\tr[\rho_A\rho_l(\lambda_i)]\tr[\rho_A\rho_r(\lambda_i)]}{\tr(\rho_A^2)^2}
        +\sum_{i,j}\lambda_i^r\lambda_j^r\frac{\tr[\rho_l(\lambda_i)\rho_l(\lambda_j)]\tr[\rho_r(\lambda_i)\rho_r(\lambda_j)]}{\tr(\rho_A^2)^2}
    \right],
\end{align}
thus in the long-range limit $r\to+\infty$ we obtain
\begin{align}
    L_S(\rho_{AB}) \sim \log\left[
        1+2\lambda_1^r c_s
    \right]
\end{align}
where $c_s$ is a constant
\begin{align}
    c_s = \frac{\tr[\rho_A\rho_l(\lambda_1)]\tr[\rho_A\rho_r(\lambda_1)]}{\tr(\rho_A^2)^2}.
\end{align}

As for the stabilizer R\'enyi entropy part $L_M(\rho_{AB})$, we first simplify it 
\begin{align}
    L_M(\rho_{AB}) &= -\log\tr(Q_2\rho_{AB}^{\otimes4}) - \log 2^2 - 2(-\log\tr(Q_1\rho_A^{\otimes4}) - \log 2)\\
    &= -\log\frac{\tr(Q_2\rho_{AB}^{\otimes 4})}{\tr(Q_1\rho_A^{\otimes4})^2}
\end{align}
and we begin from Eq.~\eqref{densitymatrix}, 
\begin{align}
    \tr(Q_2\rho_{AB}^{\otimes 4}) = \sum_{t,t'}\left[\tr(\rho_AU^t)\tr(\rho_AU^{t'})+\sum_i\lambda_i^r\tr(\rho_l(\lambda_i)U^t)\tr(\rho_r(\lambda_i)U^{t'})
    \right]^4
\end{align}
where $U^t_{s,s'} = \sigma^t_{s,s'}/\sqrt{2}$, and then
\begin{align}
    L_M(\rho_{AB}) = -\log\frac{\sum_{t,t'}\left[\tr(\rho_AU^t)\tr(\rho_AU^{t'})+\sum_i\lambda_i^r\tr(\rho_l(\lambda_i)U^t)\tr(\rho_r(\lambda_i)U^{t'})
    \right]^4}
    {\sum_{t,t'}\tr(\rho_AU^t)^4\tr(\rho_AU^{t'})^4}
\end{align}
which directly gives
\begin{align}
    L_M(\rho_{AB}) &= -\log\left[1+
    \frac{4\sum_{t,t'}(A^{tt'})^3\sum_i\lambda_i^{tt'} + 6\sum_{t,t'}(A^{tt'})^2(\sum_i\lambda_i^{tt'})^2+ 4\sum_{t,t'}A^{tt'}(\sum_i\lambda_i^{tt'})^3+ (\sum_i\lambda_i^{tt'})^4}
    {\sum_{t,t'}(A^{tt'})^4}
    \right]\nonumber \\
    & = -\log\left[
        1+4\sum_{t,t'}a_1^{tt'}\lambda^{tt'}+6\sum_{t,t'}a_2^{tt'}(\lambda^{tt'})^2 + 4 \sum_{t,t'}a_3^{tt'}(\lambda^{tt'})^3 + \sum_{t,t'}(\lambda^{tt'})^4
    \right]
\end{align}
where we use the follwing notations for simplicity
\begin{align}
    A^{tt'} &= \tr(\rho_AU^t)\tr(\rho_AU^{t'})\\
    a_k^{tt'} &= \frac{(A^{tt'})^{4-k}}{\sum_{t,t'}(A^{tt'})^4}\\
    \lambda_i^{tt'} &= \lambda_i^r\tr(\rho_l(\lambda_i)U^t)\tr(\rho_r(\lambda_i)U^{t'})\\
    \lambda^{tt'} &= \sum_i\lambda_i^{tt'}
\end{align}
and in the long-range limit $r\to+\infty$, the above expression gives
\begin{align}
    L_M(\rho_{AB})&\sim-\log[1+4\sum_{t,t'}a_1^{tt'}\lambda_1^{tt'}]\\
    &\sim -\log\left[1+4\lambda_1^r\sum_{t,t'}a_1^{tt'}
    \tr(\rho_l(\lambda_i)U^t)\tr(\rho_r(\lambda_i)U^{t'})\right]\\
    &\sim -\log[1+2\lambda_1^r c_m]
\end{align}
where $c_m = 2\sum_{t,t'}a_1^{tt'}\tr(\rho_l(\lambda_i)U^t)\tr(\rho_r(\lambda_i)U^{t'})$ is a constant.
Then the mutual SRE of any injective TIMPS in the long-range limit is given by ($0\le|\lambda_1|<1$, $r$ is the distance between two subsystems)
\begin{align}
    \lim_{r\to+\infty}L(\rho_{AB}) = \lim_{r\to+\infty}\log\left[
        \frac{1+2\lambda_1^r c_s}{1+2\lambda_1^rc_m}
    \right] = 0.
\end{align}
Here, we note that $\lambda_1$ is directly related to the correlation length $\xi$ by the follwoing relation
\begin{align}
    \xi = -\frac{1}{\ln|\lambda_1|}.
\end{align}
Other behaviour of long-range non-stabilizerness can be more accurately evaluated
\begin{align}
    L(\rho_{AB})\sim\log\left[
        \frac{
            1+2\sum_i\lambda_i^r c_i
        }
        {
            1+4\sum_i\lambda_i^r d_i
        }
    \right]
\end{align}
where 
\begin{align}
    c_i &= \frac{\tr[\rho_A\rho_l(\lambda_i)]\tr[\rho_A\rho_r(\lambda_i)]}{\tr(\rho_A^2)^2}, \\
    d_i &= \sum_{t,t'}a_1^{tt'}\tr(\rho_l(\lambda_i)U^t)\tr(\rho_r(\lambda_i)U^{t'}).
\end{align}

\section{Calculation of two-point mutual SRE in Ising model}
We consider the following transverse field Ising model (TFIM)
\begin{align}
    \hat{H}_{\mathrm{TFIM}} = -h\sum_i\hat{\sigma}_i^x - J\sum_i\hat{\sigma}^z_i\hat{\sigma}^z_{i+1} 
\end{align}
and perform local unitary transformation, 
\begin{align}
    \hat{H}_{\mathrm{TFIM}}' &= -h\sum_i \hat{H}_i\sigma_i^x\hat{H}_i - 
    J\sum_i\hat{H}_i\hat{\sigma}^z_i\hat{H}_i \hat{H}_{i+1}\hat{\sigma}^z_{i+1}\hat{H}_{i+1}\\
    &= -h\sum_i\hat{\sigma}_i^z - J\sum_i\hat{\sigma}^x_i\hat{\sigma}^x_{i+1} 
\end{align}
where $\hat{H}$ is a Hadamard gate
\begin{align}
    \hat{H}_i = \frac{\hat{\sigma}^z_i+\hat{\sigma}^x_i}{\sqrt{2}}, 
\end{align}
and it satisfies $\hat{H}^{\dagger} = \hat{H}^{-1}=\hat{H}$. 

The Hadamard gate is a Clifford operation, so the mSRE in model described by $\hat{H}_{\mathrm{TFIM}}'$ has no 
difference with the one in $\hat{H}_{\mathrm{TFIM}}$. We focus on the second Ising model $\hat{H}_{\mathrm{TFIM}}'$ in the following. 

The Hamiltonian $\hat{H}_{\mathrm{TFIM}}'$ can be exactly solved by Jordan-Wigner transformation. The Hamiltonian 
written in fermion operators is 
\begin{align}
    \hat{H}_{\mathrm{TFIM}}' = h\sum_j (2\hat{c}_j^{\dagger}\hat{c}_j-1) - 
    J\sum_j(\hat{c}_j^{\dagger}\hat{c}_{j+1} + \hat{c}_j^{\dagger}\hat{c}_{j+1}^{\dagger}+h.c.),
\end{align}
and using Fourier transformation,
\begin{align}
    \hat{c}_k &= \frac{1}{\sqrt{L}}\sum_{j}e^{-ikj}\hat{c}_j\\
    \hat{c}_j &= \frac{1}{\sqrt{L}}\sum_k e^{ikj}\hat{c}_k
\end{align}
we obtain the Hamiltonian in momentum space
\begin{align}
    \hat{H}_{\mathrm{TFIM}}' = \begin{pmatrix}
        \hat{c}_k^{\dagger} & \hat{c}_{-k}
    \end{pmatrix}\begin{pmatrix}
        2(h-J\cos k) & -i2J\sin k\\
        i2J\sin k & -2(h-J\cos k)
    \end{pmatrix}\begin{pmatrix}
        \hat{c}_k \\
        \hat{c}_{-k}^{\dagger}
    \end{pmatrix}
\end{align}
thus we can directly diagonalize the Hamiltonian and obtain the energy spectrum $\epsilon_{k\pm} = \pm\epsilon_k$
\begin{align}
    \epsilon_k = 2J\sqrt{\left(\frac{h}{J}-\cos k\right)^2+\sin^2 k}
\end{align}
with corresponding eigenvectors $(v_{k\pm},u_{k\pm})^T$
\begin{align}
    \begin{pmatrix}
        v_{k-}\\
        u_{k-}
    \end{pmatrix} &= \begin{pmatrix}
        v_k \\ u_k
    \end{pmatrix} = \frac{1}{\sqrt{2\epsilon_k(\epsilon_k+z_k)}}\begin{pmatrix}
        iy_k\\ \epsilon_k+z_k
    \end{pmatrix}\\
    \begin{pmatrix}
        v_{k+}\\
        u_{k+}
    \end{pmatrix} &= \begin{pmatrix}
        u_k^* \\ -v_k^*
    \end{pmatrix} = \frac{1}{\sqrt{2\epsilon_k(\epsilon_k+z_k)}}\begin{pmatrix}
        \epsilon_k+z_k\\ iy_k
    \end{pmatrix}
\end{align}
where we define $y_k = 2J\sin k$ and $z_k=2(h-J\cos k)$. 
Now we can explicitly write the ground state of the Ising model, 
\begin{align}
    \ket{\psi_0} = \prod_{k>0}(u_k+v_k\hat{c}_k^{\dagger}\hat{c}_{-k}^{\dagger})\ket{0}
\end{align}
where $\ket{0}$ is the vacuum state for fermions.

To calculate the two-point mSRE, we first calculate the two-point density matrix which can be 
expressed as 
\begin{align}
    \rho_{ij} = \sum_{\alpha,\beta=0}^3R_{\alpha\beta} \hat{\sigma}^{\alpha}_i\hat{\sigma}^{\beta}_j. 
\end{align}
Due to the global $\mathbb{Z}_2$ symmetry of Ising model Hamiltonian $\hat{H}_{\mathrm{TFIM}}'$, the coefficients of the 
terms like $\sigma^x_i\sigma^z_j$ which anti-commute with $\hat{F} = \prod_j \hat{\sigma}^z_j$ should be zero. 
This is because the system's symmetry requires, if there is no spontaneous symmetry breaking, 
the two-point density matrix should be invariant under the global $\mathbb{Z}_2$ transformation, which is 
\begin{align}
    \hat{F}\rho_{ij}\hat{F}^{\dagger} = \rho_{ij} \Rightarrow [\hat{F},\rho_{ij}]=0.
\end{align}
Therefore, the $R_{\alpha}$ matrix has the following form
\begin{align}
    R_{\alpha\beta} = \frac{1}{4}\begin{pmatrix}
        1 & 0 & 0 & \langle\hat{\sigma}^z_i\rangle\\
        0 & \langle\hat{\sigma}^x_i\hat{\sigma}^x_j\rangle & 0 & 0\\
        0 & 0 & \langle\hat{\sigma}^y_i\hat{\sigma}^y_j\rangle & 0\\
        \langle\hat{\sigma}^z_j\rangle & 0 & 0 & \langle\hat{\sigma}^z_i\hat{\sigma}^z_j\rangle
    \end{pmatrix}.
\end{align}
We note that there is no $\sigma^x_i\sigma^y_j$ term because the TFIM Hamiltonian is also 
invariant under time-reversal operation denoted $T=K$ and $[K,\sigma^x_i\sigma^y_j]\neq 0$ 
where $K$ is the complex conjugation operator. 

Similarly, the single-point density matrix can be expressed as
\begin{align}
    \rho_i = \frac{1}{2}\left(\mathbb{I} + \langle\hat{\sigma}^z_i\rangle\hat{\sigma}^z_i\right).
\end{align}

Now we only need to calculate the correlation functions 
$\langle\sigma^x_i\sigma^x_j\rangle,\langle\sigma^y_i\sigma^y_j\rangle,\langle\sigma^z_i\sigma^z_j\rangle$ and 
average value of $\sigma^z_i$. 
Define the one-particle Green's functions as follows
\begin{align}
    G_{jj'} &= \bra{\psi_0}\hat{c}_j\hat{c}_{j'}^{\dagger}\ket{\psi_0}, \\
    F_{jj'} &= \bra{\psi_0}\hat{c}_j\hat{c}_{j'}\ket{\psi_0}, 
\end{align}
in the thermodynamic limit, we have
\begin{align}
    G_{jj'} &= \int_0^{\pi}\frac{\mathrm{d} k}{2\pi}\frac{z_k}{\epsilon_k}\cos[k(j-j')]+\frac{1}{2}\delta_{j,j'}\\
    F_{jj'} &= \int_0^{\pi}\frac{\mathrm{d}k}{2\pi}\frac{y_k}{\epsilon_k}\sin[k(j-j')].
\end{align}

Using the above Green's functions combined with Wick theorem, we can obatin exact values of correlation functions. 
Here we direclty provide the results and for technical details, please refer to \cite{Glen2024}, 
\begin{align}
    \langle\hat{\sigma}^z_i\rangle &= 2\left(G_{i,i}-\frac12\right)\\
    \langle\hat{\sigma}^z_i\hat{\sigma}^z_j\rangle &= 4\left(G_{i,i}-\frac12\right)\left(G_{j,j}-\frac12\right)+4G_{i,j}(\delta_{i,j}-G_{j,i})+4|F_{i,j}|^2\\
    \langle\hat{\sigma}^x_i\hat{\sigma}^x_j\rangle &= \det \begin{pmatrix}
        M_{i,i+1} & M_{i,i+2} & \cdots & M_{i, j} \\
        M_{i+1,i+1} & M_{i+1,i+2} & \cdots & M_{i+1,j} \\
        \vdots & \vdots & \ddots & \vdots \\
        M_{j-2,i+1} & M_{j-2,i+2} & \cdots & M_{j-2,j} \\
        M_{j-1,i+1} & M_{j-1,i+2} & \cdots & M_{j-1,j}
    \end{pmatrix}\\
    \langle\hat{\sigma}^y_i\hat{\sigma}^y_j\rangle &= \det \begin{pmatrix}
        M_{i+1,i} & M_{i+1,i+1} & \cdots & M_{i+1, j-1} \\
        M_{i+2,i} & M_{i+2,i+1} & \cdots & M_{i+2,j-1} \\
        \vdots & \vdots & \ddots & \vdots \\
        M_{j-1,i} & M_{j-1,i+1} & \cdots & M_{j-1,j-1} \\
        M_{j,i} & M_{j,i+1} & \cdots & M_{j,j-1}
    \end{pmatrix}
\end{align}
where $M_{i,j}$ is defined as 
\begin{align}
    M_{i,j} = \delta_{i,j}-2(G_{i,j}+F_{i, j}).
\end{align}

Using the above results, we can give exact values of two-point mutual SRE in the TFIM.

\end{document}